\documentclass[aps,prb,twocolumn,groupedaddress]{revtex4-1}
\usepackage{graphicx}  % need for figures
\usepackage{hyperref}
\usepackage{color}
\usepackage{amsmath,amssymb}

\begin{document}

%\preprint{}

\title{Diffusion and transformation kinetics of small Helium clusters
  in bulk Tungsten.}


\author{Danny Perez}
\email{danny\_perez@lanl.gov}
\affiliation{Theoretical Division T-1, Los Alamos National Laboratory, Los Alamos, New Mexico 87545, USA}

\author{Thomas Vogel}
\affiliation{Theoretical Division T-1, Los Alamos National Laboratory, Los Alamos, New Mexico 87545, USA}

\author{Blas P. Uberuaga}
\affiliation{Materials Science and Technology MST-8, Los Alamos National Laboratory, Los Alamos, New Mexico 87545, USA}

%\author{Arthur F. Voter}
%\affiliation{Theoretical Division T-1, Los Alamos National Laboratory, Los Alamos, New Mexico 87545, USA}


\date{June 24, 2014}

\begin{abstract}
  The production of energy through nuclear fusion poses serious
  challenges related to the stability and performance of materials in
  extreme conditions. In particular, the constant bombardment of the
  walls of the reactor with high doses of He ions is known to lead to
  deleterous changes in their microstructures. These changes follow
  from the aggregation of He into bubbles that can grow and blister,
  potentially leading to the contamination of the plasma, or to the
  degradation of their mechanical properties.  We computationally
  study the behavior of small clusters of He atoms in W in conditions
  relevant to fusion energy production. Using a wide range of
  techniques, we investigate the thermodynamics of the clusters and
  their kinetics in terms of diffusivity, growth, and breakup, as well
  as mutation into nano-bubbles. Our study provides the essential
  ingredients to model the early stages of He exposure leading up to
  the nucleation of He bubbles.
\end{abstract}

\pacs{}
%\keywords{}

\maketitle


\section{Introduction}

One of the primary challenges in the development of fusion energy
sources is related to materials stability and performance. The
materials in a fusion reactor such as ITER experience extreme
environments, primarily the interaction with the plasma and damage by
fusion neutrons. In particular, the plasma bombards the first wall
materials with high doses of He ions which cause significant
restructuring and evolution of the material. {While the incident
  energy of the ions is modest ($\sim$50 eV), they can still penetrate
a few nm into the materials and subsequently diffuse into the bulk.}
In these plasma facing
materials, for which tungsten is a leading
candidate~\cite{Janeschitz2001,Bolt2004}, this He intake
leads to the formation of features such as He
blisters~\cite{Tokunaga2005} and so-called fuzz, a micron-thick tangle
of nm-thick tungsten
whiskers~\cite{Takamura2006,Baldwin2008,Baldwin2009,Nishijima2004}.
The formation of these features both changes how the surface of the
material interacts with the plasma but also leads to the release of
tungsten into the plasma, severely disrupting its
performance~\cite{Sharpe2002}. Thus, there is a strong imperative to
both understand and, ultimately, control the evolution of tungsten in
the presence of plasmas. This, in turn, requires a detailed
understanding of how components of the plasma, such as He, interact
with the material at the atomic scale. {Note that He is not
  generated by transmutation reactions \cite{Gilbert2012}, so that its presence in the
material can be exclusively attributed to the contact with the plasma.}

The evolution of He in metals has been studied for over 50 years and
that research has lead to a picture of He evolution that is both
complex and intriguing. He, introduced either via implantation in
laboratory experiments or via transmutation in reactor applications,
diffuses through the lattice, either encountering traps or other He
atoms, as there is significant binding between He atoms due to elastic
interactions resulting from the repulsion between the He and the metal
atoms~\cite{Becquart2012}. Once He clusters grow to a certain size,
they can force the emission of interstitials from the cluster,
creating vacancies (V) that accommodate the cluster. This so-called trap
mutation process was first identified experimentally in 1977 by van
Veen, et~al., in which they found Kr impurities in W would trap He and
lead to the formation of Frenkel pairs~\cite{vanVeen1977}. Soon after,
in 1981, Wilson, Bisson, and Baskes performed molecular dynamics
simulations of He in Ni in which they directly observed this trap
mutation process, though they dubbed it
self-trapping~\cite{Wilson1981}. As the He-V clusters accumulate more
He, forming proper bubbles, they grow via the continual emission of
interstitials and, as they grow larger, interstitial loops. This loop
punching mechanism was first described in 1959 as a mechanism by which
silver precipitates in silver halides released strain as they
grew~\cite{Parasnis1959} and discussed first in the context of He
bubbles by Greenwood, Foreman and Rimmer that same
year~\cite{Greenwood1959}. Trap mutation and loop punching formed the
basis of a model of He bubble nucleation and growth proposed by Baskes
and coworkers~\cite{Baskes1981,Baskes1983}, in which they concluded
that such mechanisms were responsible for the complex retention
behavior of He in metals. The ability of He clusters and bubbles to
grow via trap mutation-like processes has several important
implications. First, He bubbles can nucleate and grow under conditions
where vacancy mobility is essentially non-existent~\cite{Evans1981}.
Second, this growth is insaturable, meaning a ``single'' trap site can
accommodate a limitless number of He atoms~\cite{Kornelsen1980}.
Third, He thus drives the generation of vacancies and interstitials
within the material~\cite{Puska1984,Iwakiri2000,Lhuillier2013}.

Since these early studies, a large body of work, both experimental and
theoretical, has focused on the migration of He and the nucleation and
growth of small He clusters and bubbles in metals. Here, we summarize
some of the key results for He behavior in BCC metals, with a focus on
W. In He desorption experiments in Mo, Caspers et~al. \cite{Caspers1978} observed trap
mutation of He$_N$V complexes into He$_N$V$_2$ complexes for
$N>6$. Lhuillier et~al. used NRA (Nuclear Reaction
Analysis) and PAS (Positron Annihilation Spectroscopy) to show that He
bubble formation in W does not require preexisting vacancies but can
form via trap mutation~\cite{Lhuillier2013}, though radiation-induced
vacancies can also serve as nucleation sites for
bubbles~\cite{Iwakiri2000}. Further experimental studies have shown
that impurities can trap He in W and ultimately lead to trap mutation
processes which nucleate
bubbles~\cite{vanVeen1977,AbdElKeriem1993,Iwakiri2000}. Loop punching
from He bubbles has also been observed in experiments on
Mo~\cite{Evans1981} and W~\cite{Iwakiri2000}.

Atomistic simulations, both using electronic structure calculations
and classical potentials, have provided details into these mechanisms.
Density functional theory calculations of He in BCC metals have
focused on the properties of single He interstitials and small He
clusters and complexes. These have identified that He interstitials
reside in tetrahedral interstices in Fe and
W~\cite{Becquart2007,Becquart2012}, have very small migration energies
(of about 0.06 eV in W)~\cite{Becquart2007,Becquart2012}, and bind
strongly with impurities in Fe~\cite{Yan2011, Hao2012}. Simulations
using potentials have examined larger agglomerates of He, including
the behavior of He bubbles, within metallic matrices. They have
confirmed that trap mutation is a mechanism for He bubble nucleation
in Fe~\cite{Seletskaia2006,Stewart2010,Stewart2011,Yang2013} and
W~\cite{Henriksson2006,Sefta2013} and that loop punching is indeed a
mechanism for He bubble growth in Fe~\cite{Caro2011,Gao2011,Yang2013},
Mo~\cite{Zhang2011} and W~\cite{Henriksson2006,Sefta2013,Boisse2014}. Simulations
have also shown that H aids in the loop punching of He
bubbles~\cite{Hayward2012}. Many of these observations extend to other
metals as well, as reviewed by Trinkaus~\cite{Trinkaus1982}.

These atomistic simulations provide the detail necessary to
parameterize higher level models of He bubble evolution within
materials, which can be applied in kinetic Monte
Carlo~\cite{Morishita2007,Becquart2010,Guo2013}, rate
theory~\cite{Baskes1983,Xu2010}, and cluster
dynamics~\cite{Marian2012}, for example. However, typically, not all
of the relevant thermodynamic and kinetic properties are available,
especially regarding the rates at which He interstitial clusters
diffuse and transform into other complexes. While migration energies
and prefactors have been determined for He interstitial clusters in
Fe~\cite{Stewart2011}, systematic studies of He interstitial cluster
behavior have not been reported for He in W. The goal of the present
work is to determine the relevant thermodynamic and kinetic parameters
that describe He in W to inform higher level models. {The current
  study is concerned with the behavior of He in bulk W, an environment
that is typical of the nucleation and initial evolution of extended He
defects, as the interaction of small clusters with the surface is 
very short ranged \cite{Hu2014}. }

The paper is organized as follows: the different simulations
techniques used in this work are described in Sec.~\ref{sec:methods},
the structure and thermodynamics of He clusters are discussed in
Sec.~\ref{subsec:struct-thermo}, while their diffusion, breakup, and
mutation kinetics are investigated in Secs.~\ref{subsec:diffusion},
\ref{subsec:breakup} and~\ref{subsec:mutation}, respectively; finally,
implications for the parameterization of cluster dynamics models are
discussed in Sec.~\ref{sec:discussion}, before concluding.

\section{Methods}

\label{sec:methods}

Simulations were carried out using different atomistic simulations
methodologies, namely conventional Molecular Dynamics (MD),
Temperature Accelerated Dynamics (TAD)~\cite{sorensen2000}, and
statistical temperature (STMD)~\cite{kim06prl} and multicanonical
Molecular Dynamics~\cite{hansmann96cpl,junghans14tbp}. TAD is an
accelerated MD technique~\cite{Perez2009} that allows for an extension
of the timescale ammenable to simulations in cases where the dynamics
are activated, i.e., where evolution occurs through a sequence of rare
structural transitions separated by relatively long periods of
strictly vibrational motion.  TAD proceeds by running simulations at
elevated temperatures to speedup the occurrence of rare transitions,
and by filtering these possible transitions to select those that are
statistically appropriate for evolution at a lower temperature. The
end result is that the dynamics can be significantly accelerated when
barriers are sufficiently high. For details, the readers are refered
to the original publication~\cite{sorensen2000}. A useful side-effect
of TAD is that all possible transition are fully analyzed by the means
of a Nudged Elastic Band (NEB)~\cite{henkelman2000}.  This in turn
enables one to easily characterize available transition pathways. A
subset of the MD simulations were also fully analyzed using the NEB
method, i.e., every 100\,fs, the trajectory was interupted to determine
whether a transition occurred.  If it did, a NEB was used to identify
the pathway connecting the previous and current states. In that case,
all states and saddles points were also saved and analyzed.

The thermodynamics of this system were investigated using
multicanonical MD~\cite{hansmann96cpl,junghans14tbp}, a
generalyzed-ensemble method where one aims at performing a random walk
in a collective variable, in our case the internal energy $E$. In
practice, this is done by reweighting the interatomic forces during a
conventional canonical simulation at a reference (thermostat)
temperature $T_0$ via
\begin{equation}
f_\textrm{muca}=\frac{T_0}{T(E)}f_\textrm{can}\,.
\label{eq:muca_f}
\end{equation}
Here $T^{-1}(E)=\partial{S(E)}/\partial{E}$, with $S(E)$ being the
microcanonical entropy which is related to the density of states
$g(E)$. $T(E)$ is iteratively obtained via the STMD
approach~\cite{kim06prl}. In that scheme, one begins with a constant
initial 'guess' $T_{t=0}(E)=T_0$ (corresponding to conventional
canonical MD at $T_0$) and updates the estimator $T_t(E)$ at every
time $t$ by effectively accumulating a bias potential, as also done in
metadynamics, cf. Ref. ~\onlinecite{junghans14tbp}.  Once $T_t(E)$ is
converged, we record histograms $H(E,Q)$ of the internal energy and
other observables of interest $Q$. Those histograms can then be
reweighted to any other ensemble and observable averages can be
calculated at any temperature in the range initially covered by the
estimator for $T(E)$.

Calculations were carried out using an Embedded Atom Method (EAM)
description of the interatomic interactions, with W--W interactions
from Ackland and Thetford~\cite{Ackland1987} and modified by Juslin
and Wirth~\cite{Juslin2013}, He--He interactions from
Beck~\cite{Beck1968a,*Beck1968b} and modified by Morishita
et~al.~\cite{Morishita2003}, and He--W interactions from Juslin and
Wirth~\cite{Juslin2013}. The simulation cells contained $6\times 6
\times 6$ W unit cells, for a total of 432 W atoms. The small size of
the cell is instrumental in reaching the long timescales necessary to
characterize the dynamics at low temperatures, or for high barrier
events. {We verified that this size was adequate to properly
  capture the energetics of the clusters. For example, for $N=6$, we
  only observe a $0.3\%$ decrease in the energy of a cluster (compared to
  the perfect W bulk) by increasing the cell size by a factor of 2 in
  every direction.}
 He atoms are added as appropriate. We investigated He$_N$
clusters from $N=2$ to $7$ at temperatures covering the whole range of
relevance to fusion applications, which is centered around 1000\,K.
All He atoms are interstitial at the beginning of the \hbox{simulations}.

\section{Results}

We are interested in three primary aspects of the kinetics of He clusters in tungsten:
diffusion, breakup into smaller clusters, and ``trap mutation'' (i.e.,
conversion from an interstitial to a substitutional cluster through
the creation of a W Frenkel pair). These three mechanisms control the 
microstructural impact of He on the tungsten wall.
Each will be discussed in turn in the following sections.

\subsection{Structure and Thermodynamics}
\label{subsec:struct-thermo}

We first used TAD to identify the dominant low-temperature pathways
(i.e., the ones with the lowest energy barrier). This was done by
running TAD with a relatively low target temperature of 300\,K (in
order to maximize the likelihood of finding the lowest-barrier
pathway) and a high temperature of 600\,K. The duration of each
simulation varied, being a function of the typical barriers in the
system, but ranged from tens of nanoseconds to tens of microseconds.
The simulation cell for each size cluster started with the lowest
energy structure of the next lowest size cluster and one addition He
interstitial. The trajectory was allowed to evolve until the two
species encountered one another, forming a larger interstitial He
cluster, and further evolved to allow for diffusion of the cluster.
The trajectory was then analyzed to identify both the lowest energy
diffusion pathway for each cluster as well as the lowest energy
structure for each cluster.

\begin{figure*}[t]
 \centering
 \includegraphics[width=15. cm]{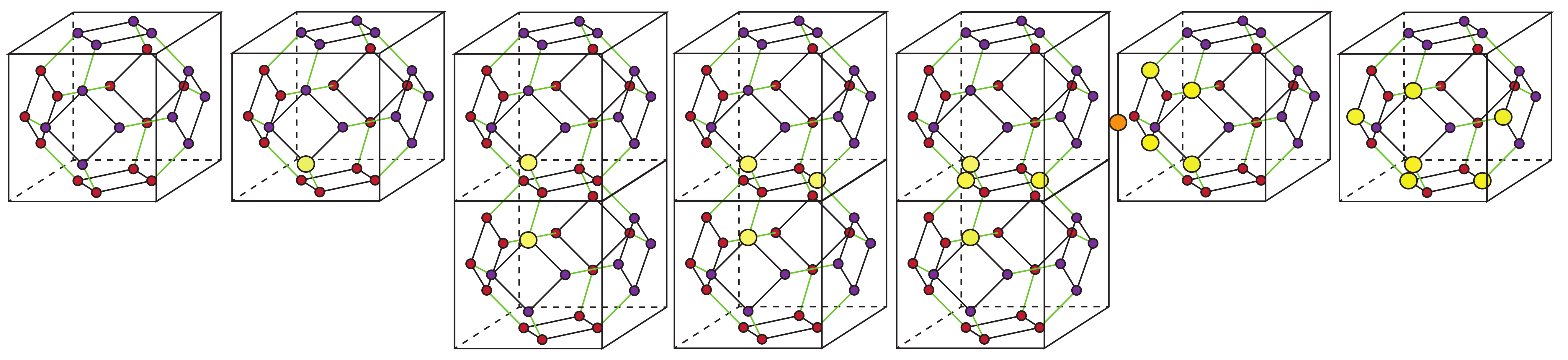}
 \caption{The lowest energy structure of each cluster for sizes $N=1$
   to 6. (a) Position of the tetrahedral interstitices within the unit
   cell of the BCC tungsten lattice, adapted from
   Ref.~\onlinecite{Szabo86}. The colors are simply guides for the eye
   to distinguish tetrahedral sites on the back faces of the unit cell
   from the front faces. The black lines connect tetrahedral sites
   separated by ${1/4,1/4,0}$ (in units of the lattice constant of the
   material) on a given face of the unit cell while the green lines
   connect tetrahedral sites separated by ${1/4,1/4,0}$ on opposing
   faces of the unit cell. Tungsten atoms, not shown, reside on the
   corners and at the body center of the cube. (b-g) Structures of He
   interstitial clusters from size 1 to 6. The position of each He
   atom within the cluster is highlighted in yellow when it occupies a
   tetrahedral site and orange (only for $N=5$) when it occupies an
   octahedral site. These are idealized positions within the
   tetrahedral interstice sublattice. In reality, atomic forces result
   in some distortion away from these ideal positions upon
   minimization of the forces.}
    \label{fig:structure}
\end{figure*} 
Figure~\ref{fig:structure} shows the structure of the ground state
cluster geometry for clusters of size $N=1$ to 6. The structure of the
tetrahedral interstices in BCC tungsten are illustrated in
Fig.~\ref{fig:structure}a while the idealized structures for the
clusters are given in Figs.~\ref{fig:structure}b-g. There are 24
tetrahedral interstices at ${1/2,1/4,0}$ positions within the BCC unit
cell, including those on periodic faces. Octahedral interstices, not
shown, reside at ${1/2,0,0}$ and ${1/2,1/2,0}$; there are 18 such
sites. For nearly all clusters, the ground state structure involves He
atoms sitting on tetrahedral interstices within the BCC lattice. The
exception is $N=5$, which is discussed below. For the other size
clusters, He atoms first arrange themselves along a $[100]$ direction
across the edge of one unit cell ($N=2$). The structure for $N=4$ is
two of these pairs lying perpendicular to one another while $N=3$
represents an intermediate structure between $N=2$ and $N=4$. The
structure for $N=6$ becomes more complex. In this case, the He atoms
all lie within the unit cell. There are still two pairs of $[100]$
oriented He atoms, but they no longer lie directly opposite one
another, as for $N=4$. This is to accommodate the other two He atoms
that lie on opposite faces of the unit cell and tend to create a more
open structure than the more compact $N=4$ structure. As mentioned,
$N=5$ is the exception in that not all of the He atoms occupy
tetrahedral positions. First, four He atoms reside on tetrahedral
sites, but break the motif of $N=4$, forming two $[100]$ pairs which
are now oriented parallel to one another. Further, the fifth He atom
occupies an octrahedral rather than a tetrahedral interstice. This
seems to have consequences for the mobility of this cluster, as will
be discussed below.

% As the size of the cluster increases, so does the binding of the cluster, as shown in Fig.~\ref{fig:energies} for cluster sizes $N=1$ to 8. Here, the binding energy is calculated as the difference in energy between the clustered He and the equivalent number of He in isolated interstitial positions at infinity. Thus, negative numbers indicated stronger binding. Coincident with the change in structural motif discussed above for $N=5$, there is a bend in the binding energy curve at that size. Larger clusters are still more strongly bound than smaller clusters, but the energy gain per He atom is less than for the smaller clusters. While data is presented for sizes up to 8 He atoms, we limit our discussion to sizes up to 6 as larger sizes, as will be discussed below, are not stable within the W lattice. 

\begin{figure}[t]
 \centering
 \includegraphics[width=7cm,trim=0.5cm 0.7cm 1.5cm 0.6cm,clip]{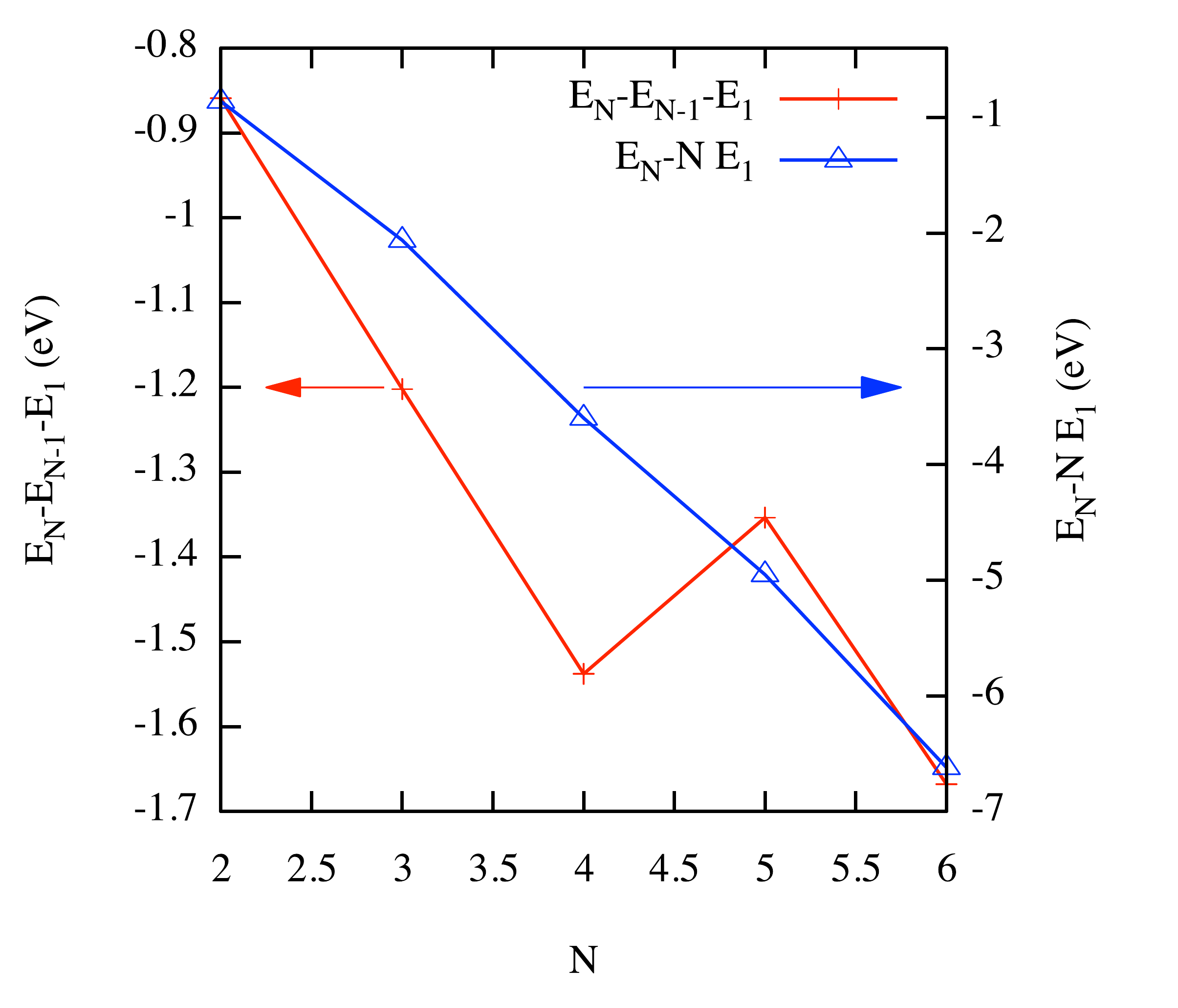}
 \caption{Energy change (at $T=0$\,K) upon removal of single atoms from
   a cluster of size $N$ ($E_N-E_{N-1}-E_1$; red, left axis) and upon
   complete fragmentation into $N$ single atoms ($E_N-NE_1$; blue,
   right axis).  }
    \label{fig:energies}
\end{figure}

As the size of the cluster increases, so does the strength of the
binding of the cluster, as shown in Fig.~\ref{fig:energies} for
cluster sizes $N=2$ to 6.  Coincident with the change in structural
motif discussed above for $N=5$, there is a bend in the $E_N-NE_1$
curve ($E_s$ being the energy of a cluster of size $s$) at that size
and a local maximum of $E_N-E_{N-1}-E_1$.

\begin{figure}[t]
 \centering
 \includegraphics[width=6.7cm,trim=0.5cm 1.7cm 0.9cm 0.6cm,clip]{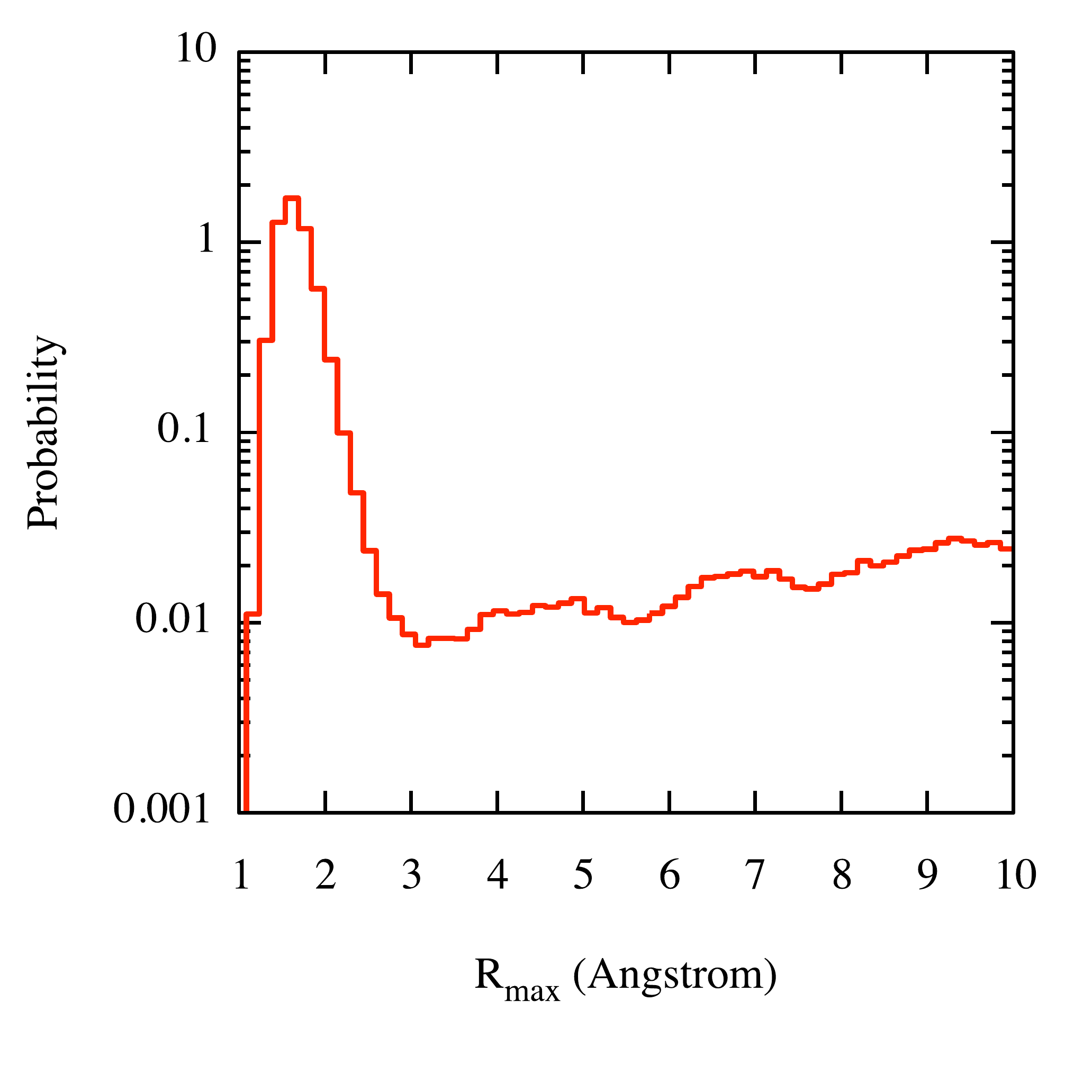}
 \caption{Probability distribution function of the He--He distance for
   $N=2$ at 1500\,K. }
 \label{fig:r-histogram}
\end{figure}

\begin{figure}
 \centering
 \includegraphics[width=.8\columnwidth,trim=0.5cm 0.6cm 1.0cm 0.6cm,clip]{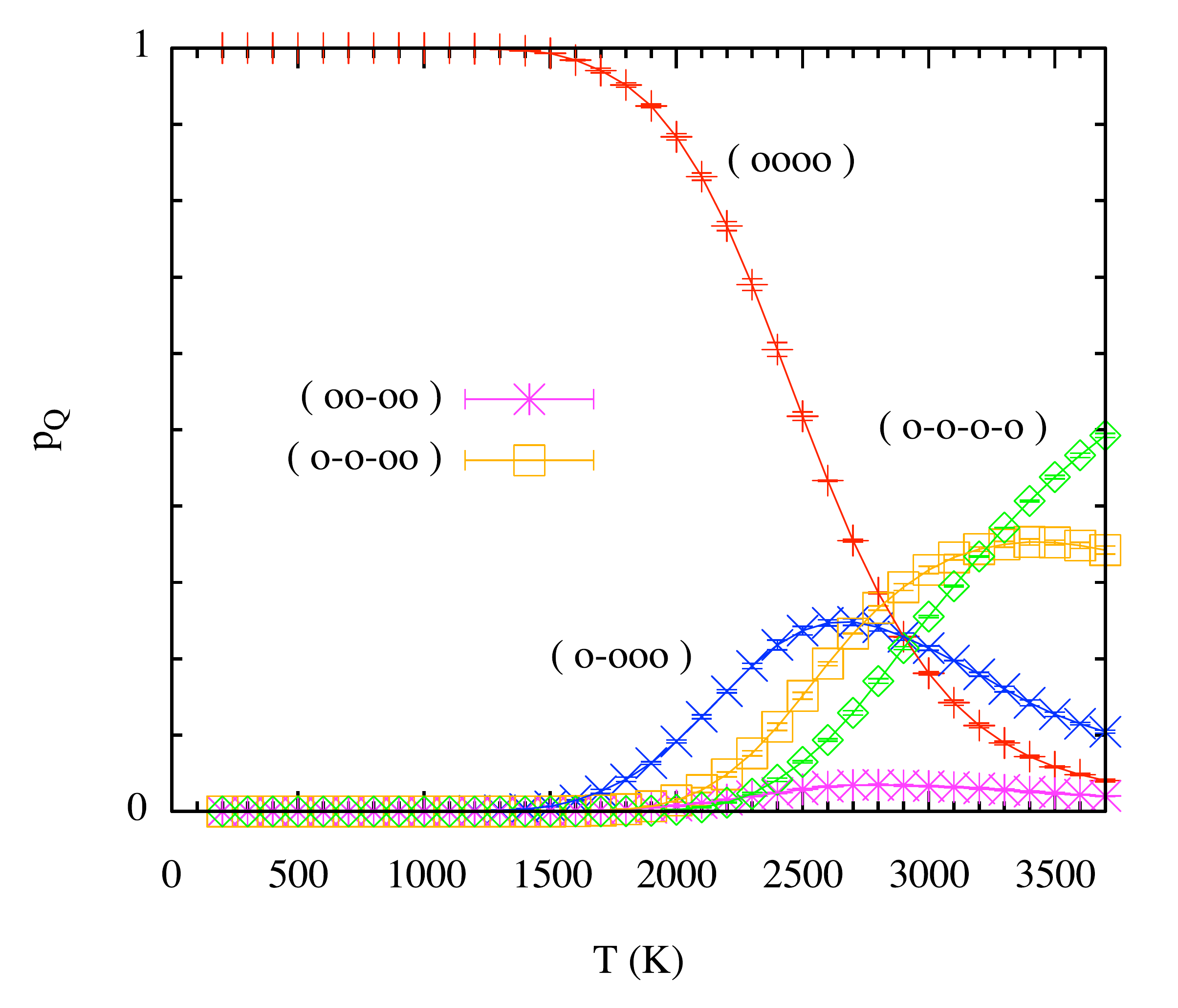}
 \caption{Probability of finding a certain cluster distribution for
   $N=4$. See text for details.}
 \label{fig:cluster_probs_4}
\end{figure}

Finite temperatures could affect the zero-temperature properties
described above, which could be particularly important for fusion
applications as operating temperatures are very high. We therefore
computed free energy differences between different clusters of sizes
$N=2$ to~6. For this purpose, two He atoms are considered to be part
of the same cluster if they are closer than a cutoff distance of
{2.8\,\AA}. This is a proper choice, as it contains most of the first
peak of the probability distribution function for He--He distance, as
shown in Fig.~\ref{fig:r-histogram} for $N=2$ and $T=1500$\,K. We verified
that the results are not significantly affected by the precise choice
of the cutoff radius.
After obtaining the proper simulation weights $w_\textrm{muca}(E)$ via
individual STMD runs, we perform a multicanonical production run
(applying Eq.~\ref{eq:muca_f}) for every $N$ and measure the
2-dimensional joint distribution of $E$ and cluster composition~$Q$.
For $N=2$ there can be $2$ different compositions (either a cluster
consisting of 2 atoms or two single He atoms), for $N=3$ there are 3
possible compositions -- a cluster of 3 (ooo), a cluster of 2 plus a
single atom (o-oo), and three single atoms (\hbox{o-o-o}), etc. By
reweighting the multicanonical histograms $H(E,Q)$ we can compute the
canonical distributions of cluster compositions at all temperatures
$T$ via
\begin{align}
P^\textrm{can}_T(Q)=\sum_E\,w_\textrm{muca}^{-1}(E)H(E,Q)\,\mathrm{e}^{-E/k_\textrm{B}T}\,,
\end{align}
where the sum extends over all energy bins.  From these distributions,
one calculates the probabilities $p_Q(T)$ to find certain cluster
compositions $Q$ at a given temperature. As an example, we plot these
probabilities for $N=4$ in Fig.~\ref{fig:cluster_probs_4}. For
temperatures $T\lesssim2500$\,K, the most prominent cluster
configuration is a single cluster containing all four He atoms. Single
atoms start to split off at $T\approx1500$\,K, though, and most
cluster compositions are present by $T\approx3000$\,K.

\begin{figure*}
 \centering
 \includegraphics[width=7cm]{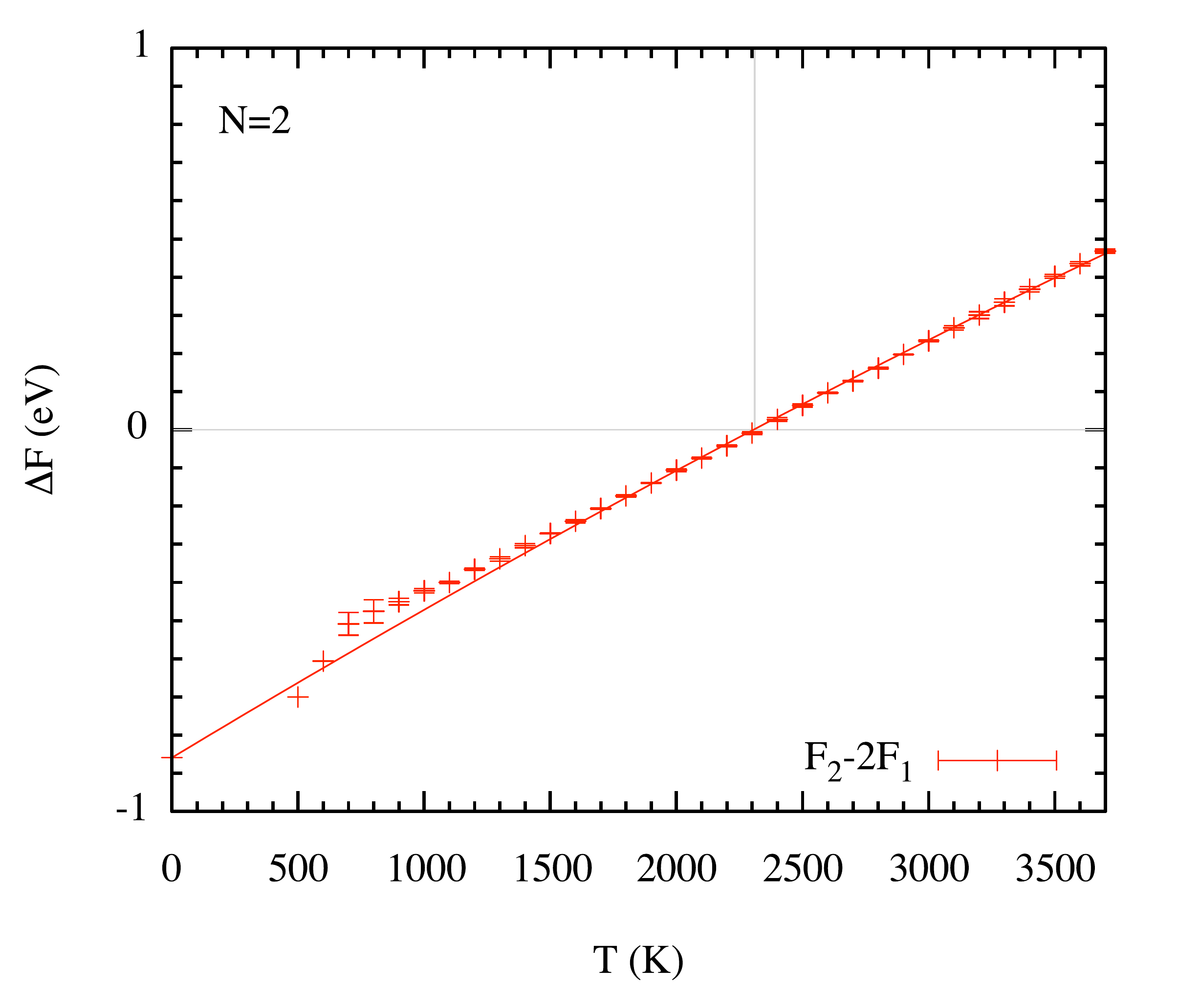}
 \includegraphics[width=7cm]{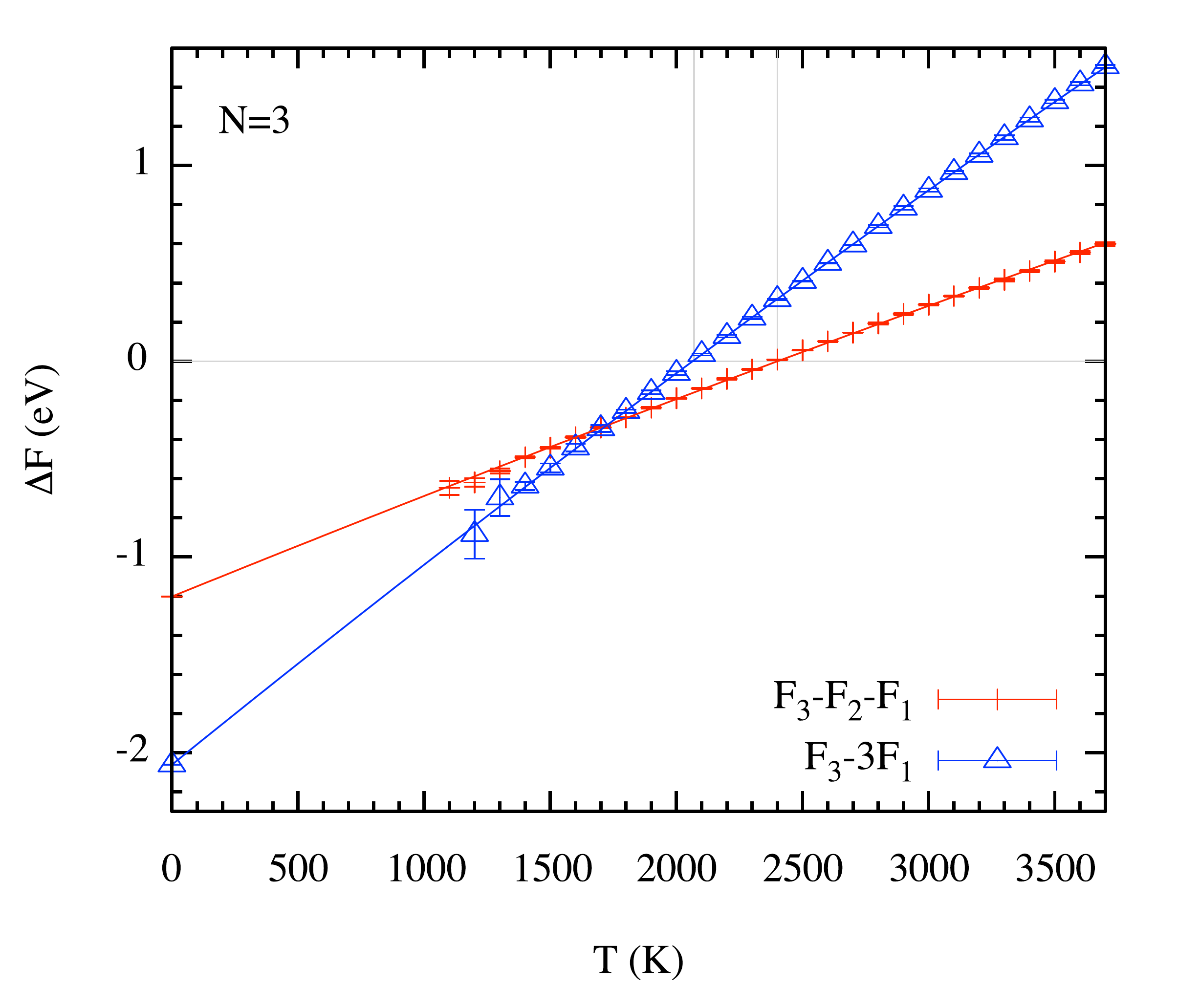}
 \includegraphics[width=7cm]{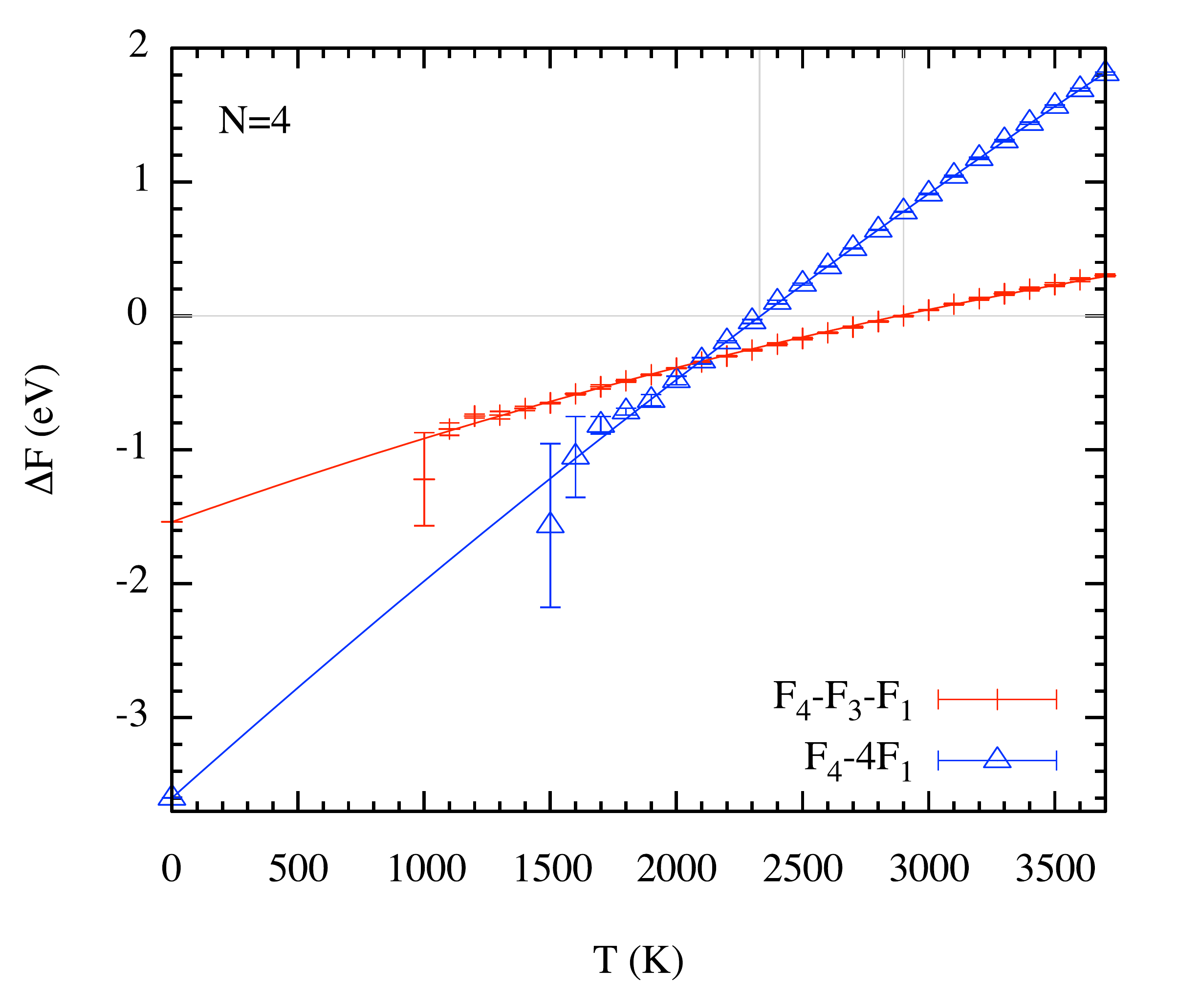}
 \includegraphics[width=7cm]{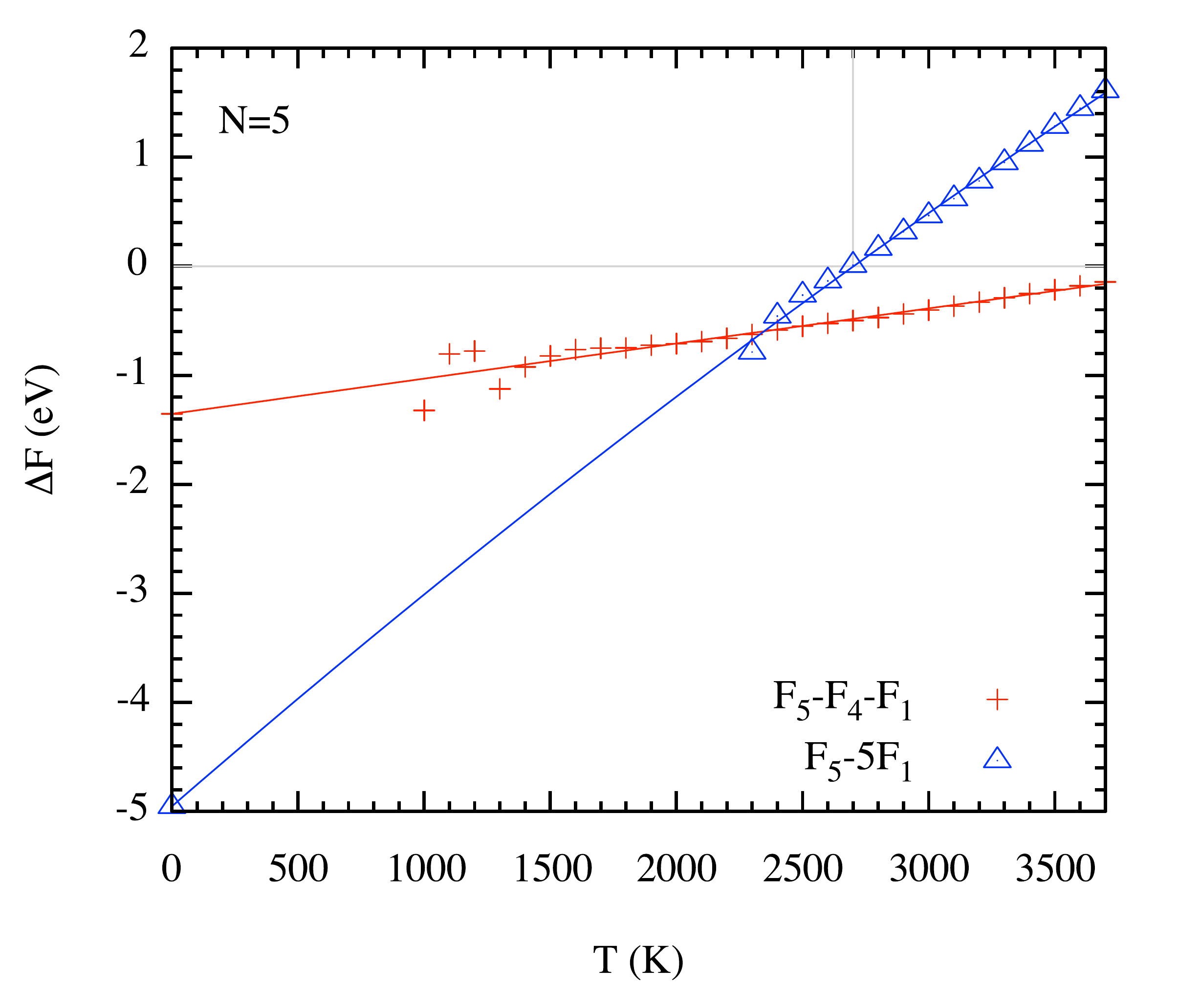}
 \includegraphics[width=7cm]{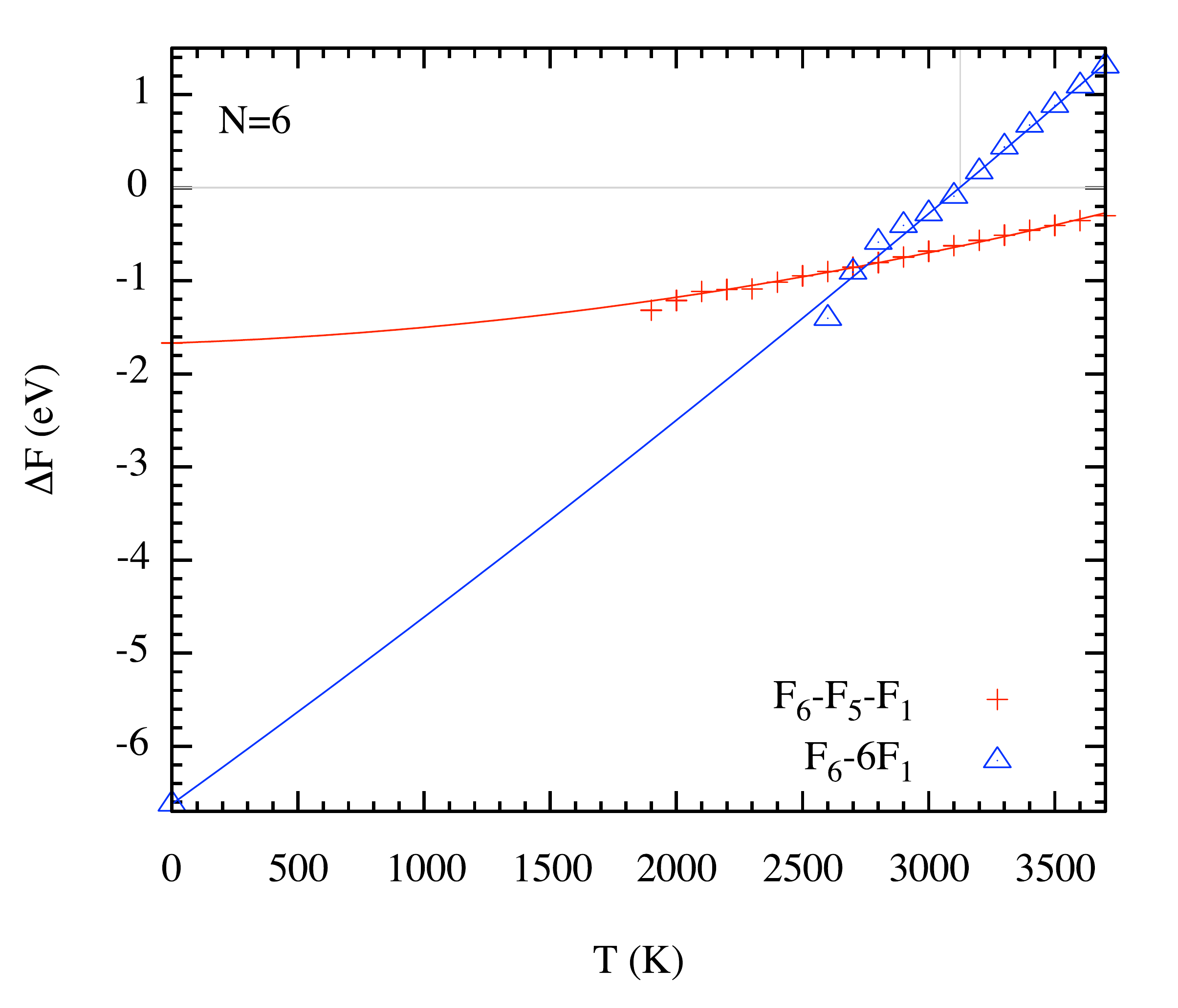}
 \caption{Free energy change upon removal of single atoms
   from a cluster of size $N$ ($F_N-F_{N-1}-F_1$; red) and upon
   complete fragmentation into $N$ single atoms ($F_N-NF_1$; blue).}
 \label{fig:free_e_all_N}
\end{figure*}

While these probabilities depend on the volume of the simulation cell,
they can be used to obtain the (volume-independent) free energy of the
different clusters.  Writing the free energy of a cluster of size $s$
as $F_s(T)=-k_\textrm{B}T\ln q_s$, with $q_s$ being the partition
function of that cluster, the free energy difference for the complete
breakup of a cluster of size $s$ into $s$ single atoms reads
$F_s-sF_1=-k_\textrm{B}T\ln(q_s/q_1^s)$.  Analogously, the free energy
change upon the loss of a single atom from a cluster of size $s$ reads
$F_s-F_{s-1}-F_1=-k_\textrm{B}T\ln(q_s/q_{s-1}q_1)$.  Based on the
formalism developed in Ref.~\onlinecite{kindt13jctc}, one can show
that the above ratios of cluster partition functions are related to
ratios of the probabilities $p_Q(T)$ to find certain cluster
compositions. For example,
$q_3/q_2q_1=p_{\textrm{(ooo)}}/p_{\textrm{(o-oo)}}$, and
$q_3/q_1^3=p_{\textrm{(ooo)}}/(3!\,p_{\textrm{(o-o-o)}})$, and so on.
The corresponding free energy differences are reported in
Fig.~\ref{fig:free_e_all_N}.
 
At low temperatures, the probabilities of finding single atoms are
vanishingly small (see Fig.~\ref{fig:cluster_probs_4}, for example),
hence it is almost impossible to accurately calculate the probability
ratios.  However, the value at $T=0$\,K can be infered from the
binding energies reported above. The lines in
Fig.~\ref{fig:free_e_all_N} are fits to the data including those
points. We note that even without including those points in the fits,
the extrapolation is almost perfect in all cases (not shown). It is
therefore unlikely that abrupt changes in the behavior of the free
energy occur below 1000\,K.  It can be qualitatively seen that the
temperature at which atoms split off the cluster or at which clusters
break up completely grows with $N$, in accordance with the behavior
observed at $T=0$\,K.
%In fact, for $N\geq5$,
%splitting off an atom from the cluster is not favorable anymore even
%for the highest temperatures close to the W melting point ($\Delta
%F<0$ for all $T$, red curves).

\subsection{Diffusion}
\label{subsec:diffusion}
A crucial characteristics of He clusters is how fast they diffuse
through the lattice.  This quantity controls the timescale on which
clusters and bubbles can grow and how far He can penetrate into bulk
W. In the following, we discuss the nature of the diffusion pathways
for clusters of sizes $N=1$ to 6, and then discuss the diffusivity and
its temperature dependence, as measured through direct MD simulation.

\subsubsection{Diffusion pathways}

\begin{figure*}[]
 \includegraphics[width=.33\textwidth]{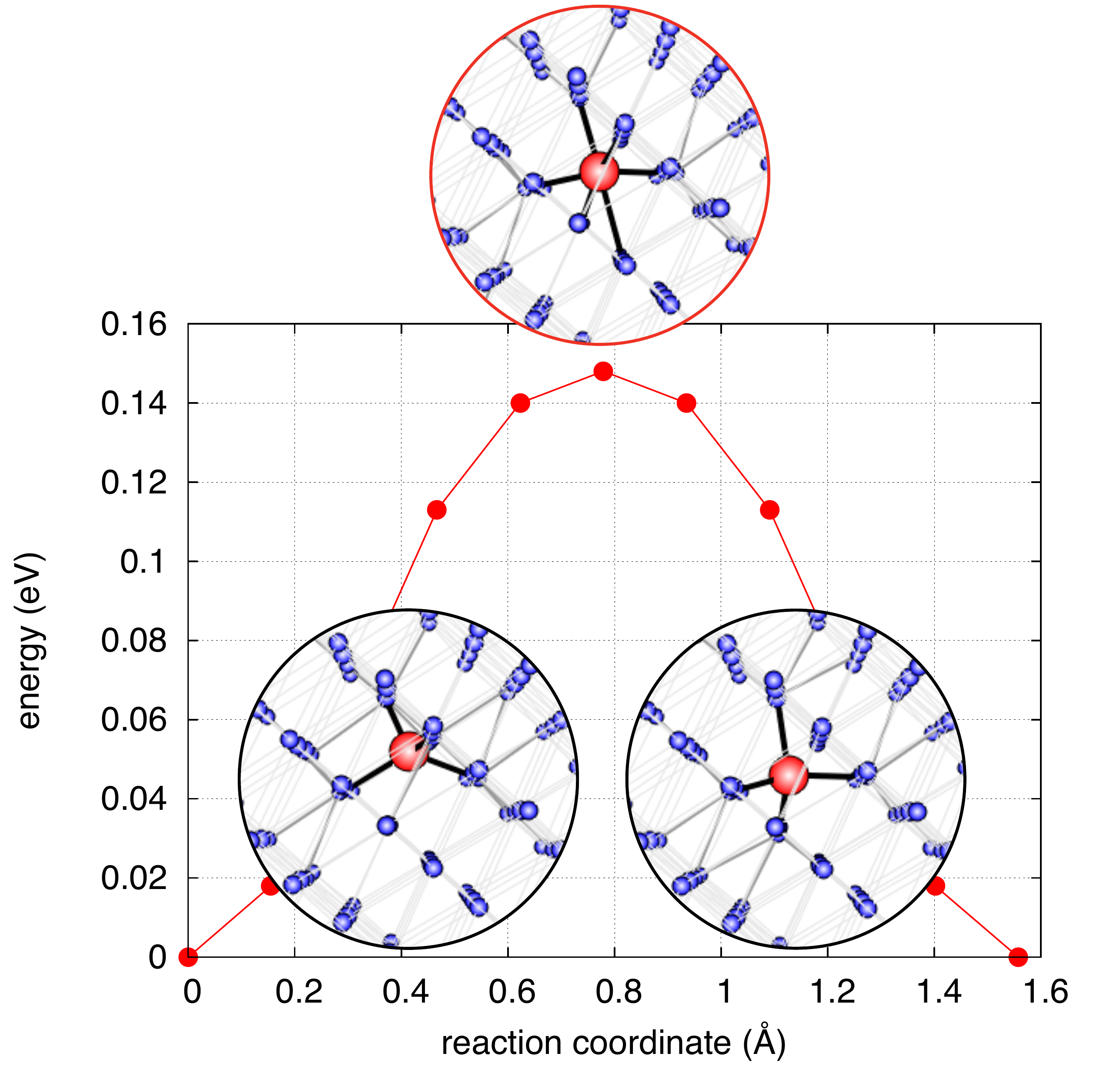}\hfill
 \includegraphics[width=.33\textwidth]{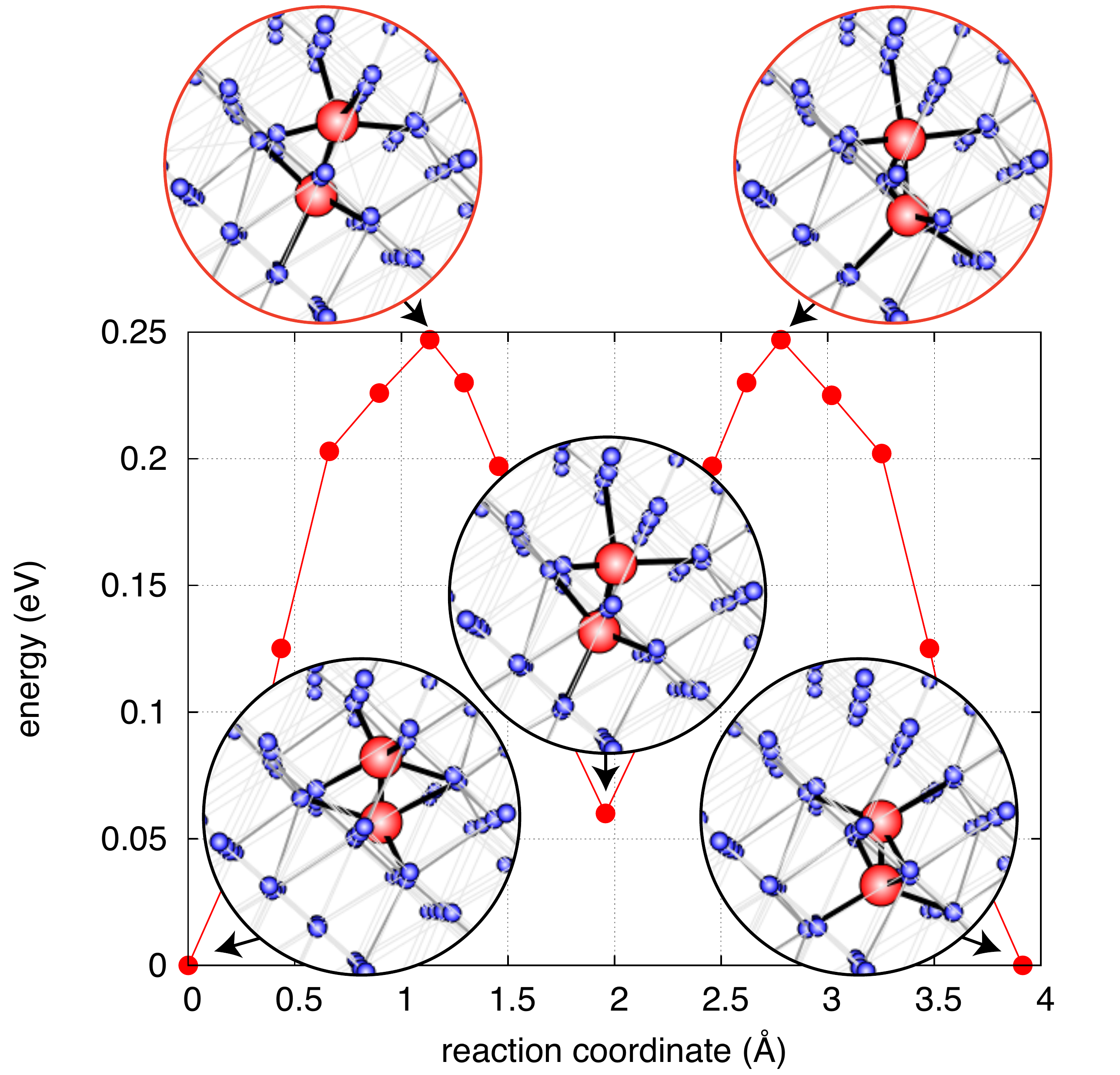}\hfill
 \includegraphics[width=.33\textwidth]{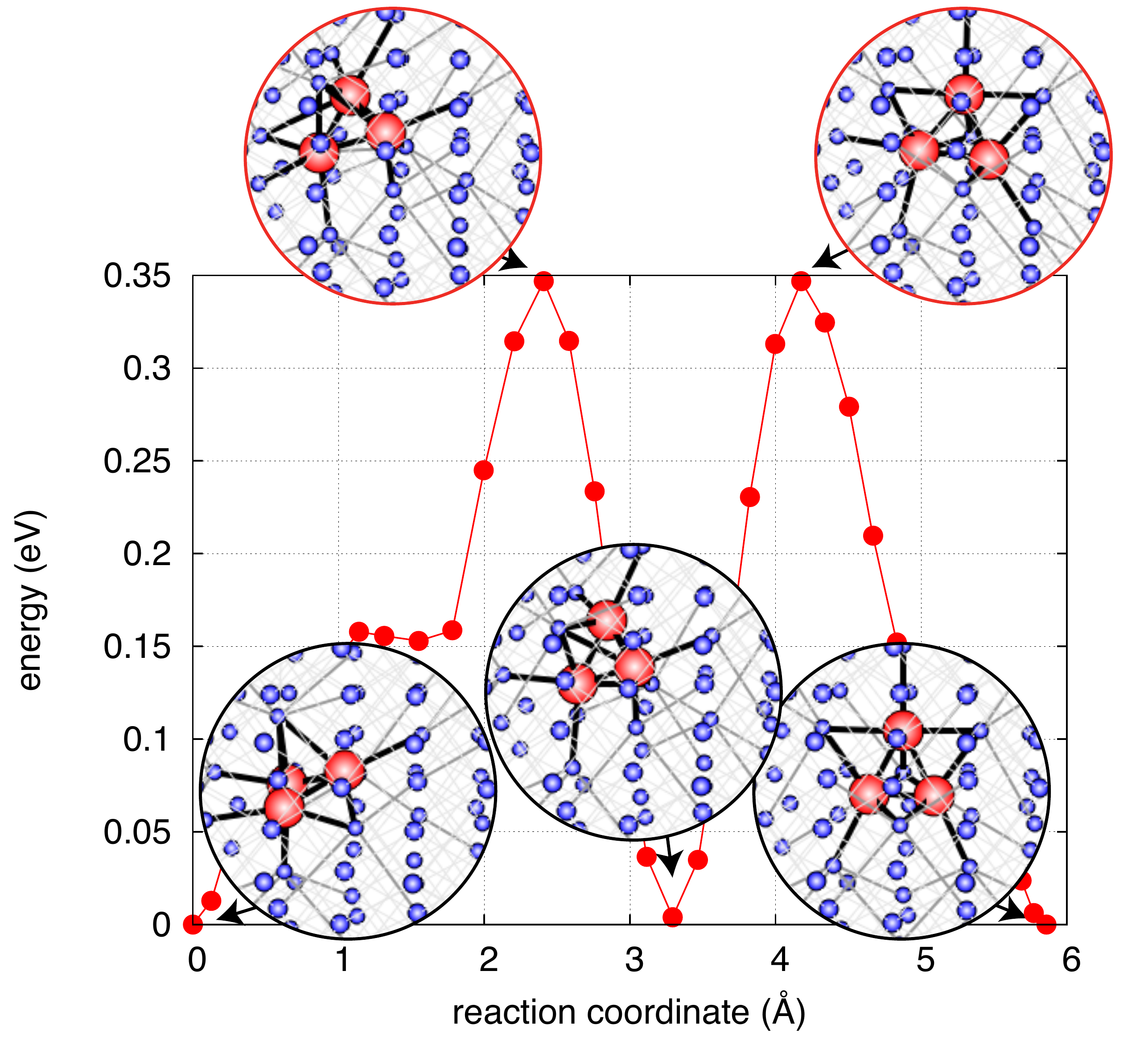}\\
 \includegraphics[width=.33\textwidth]{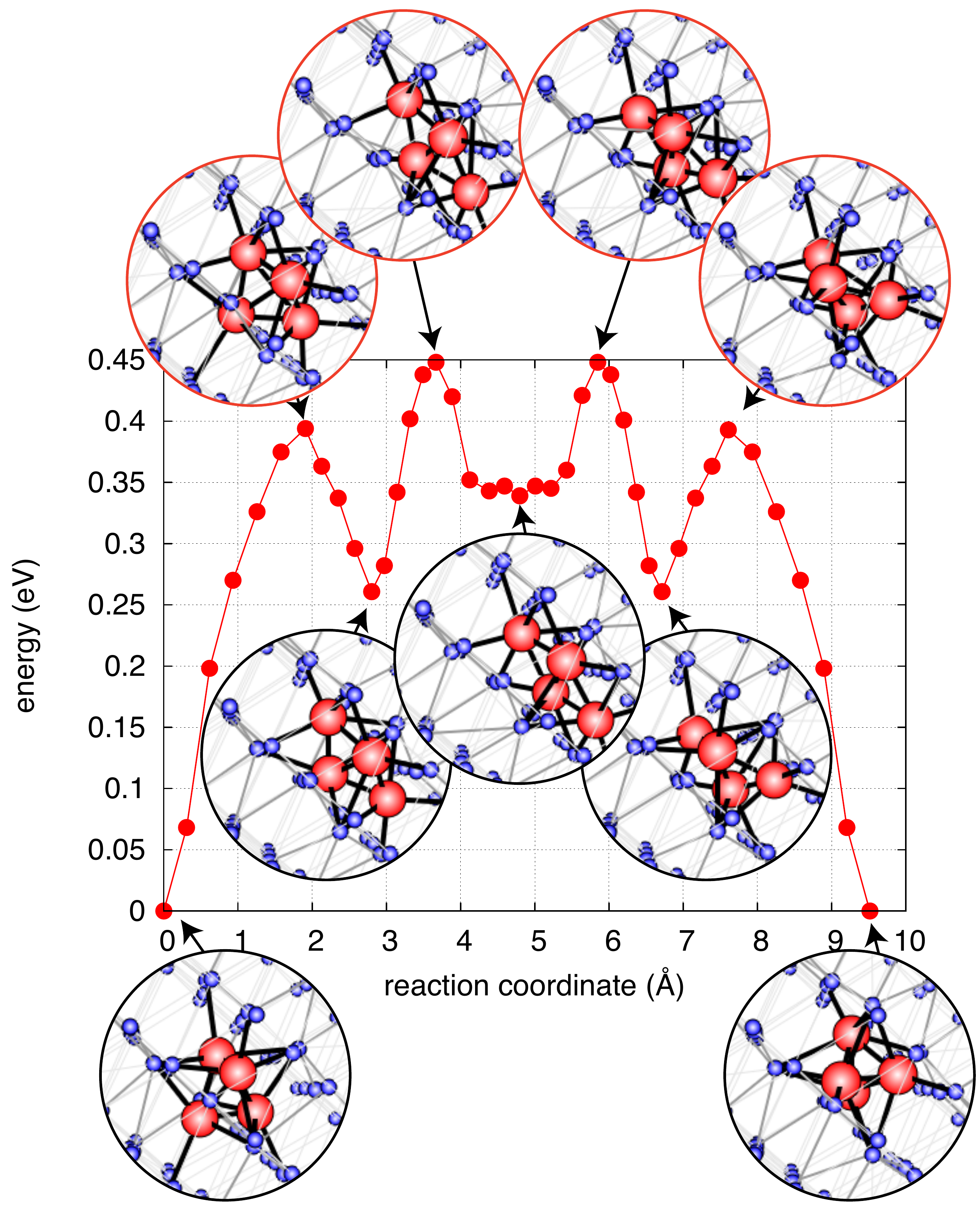}\hfill
 \includegraphics[width=.33\textwidth]{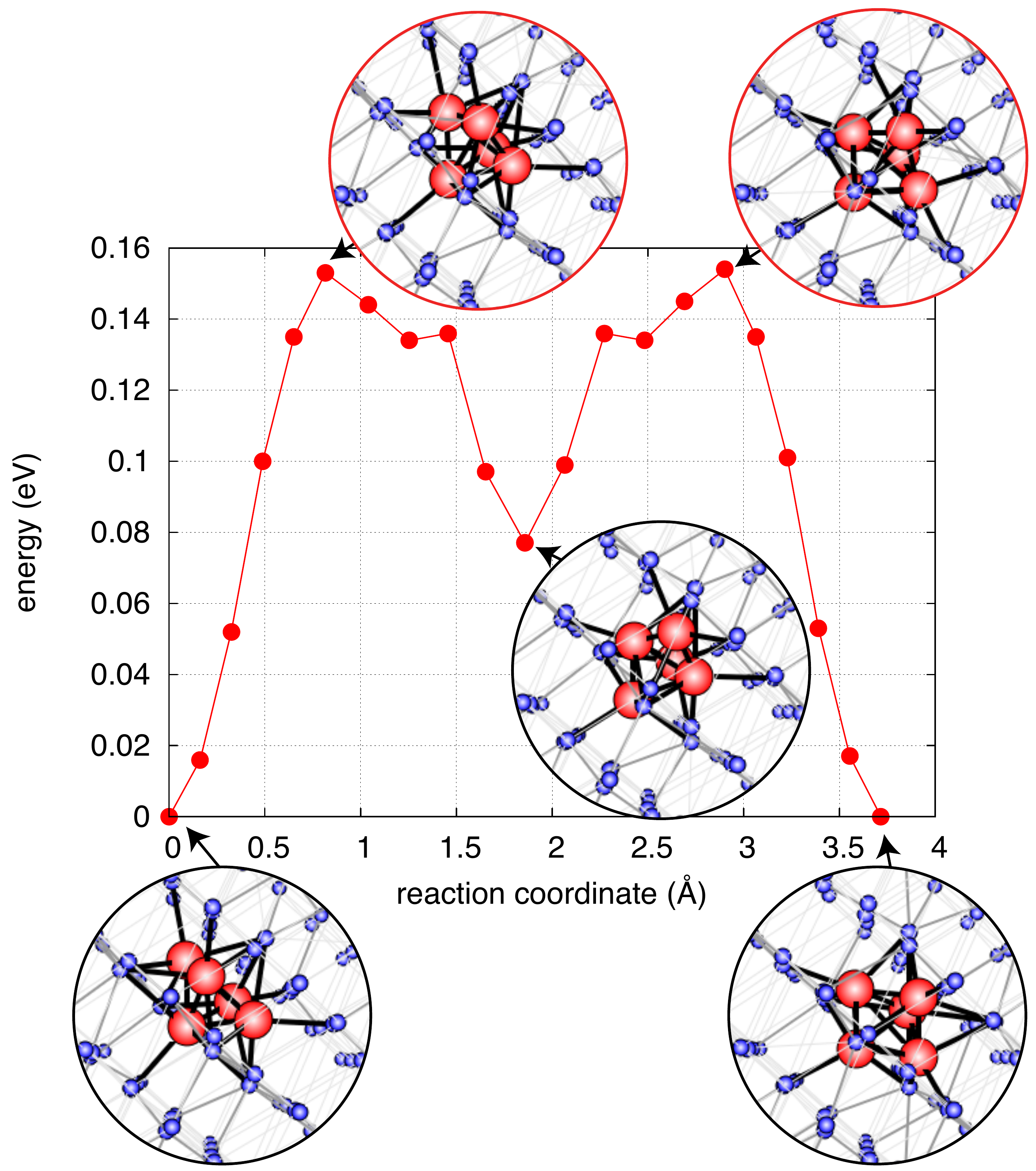}\hfill
 \includegraphics[width=.33\textwidth]{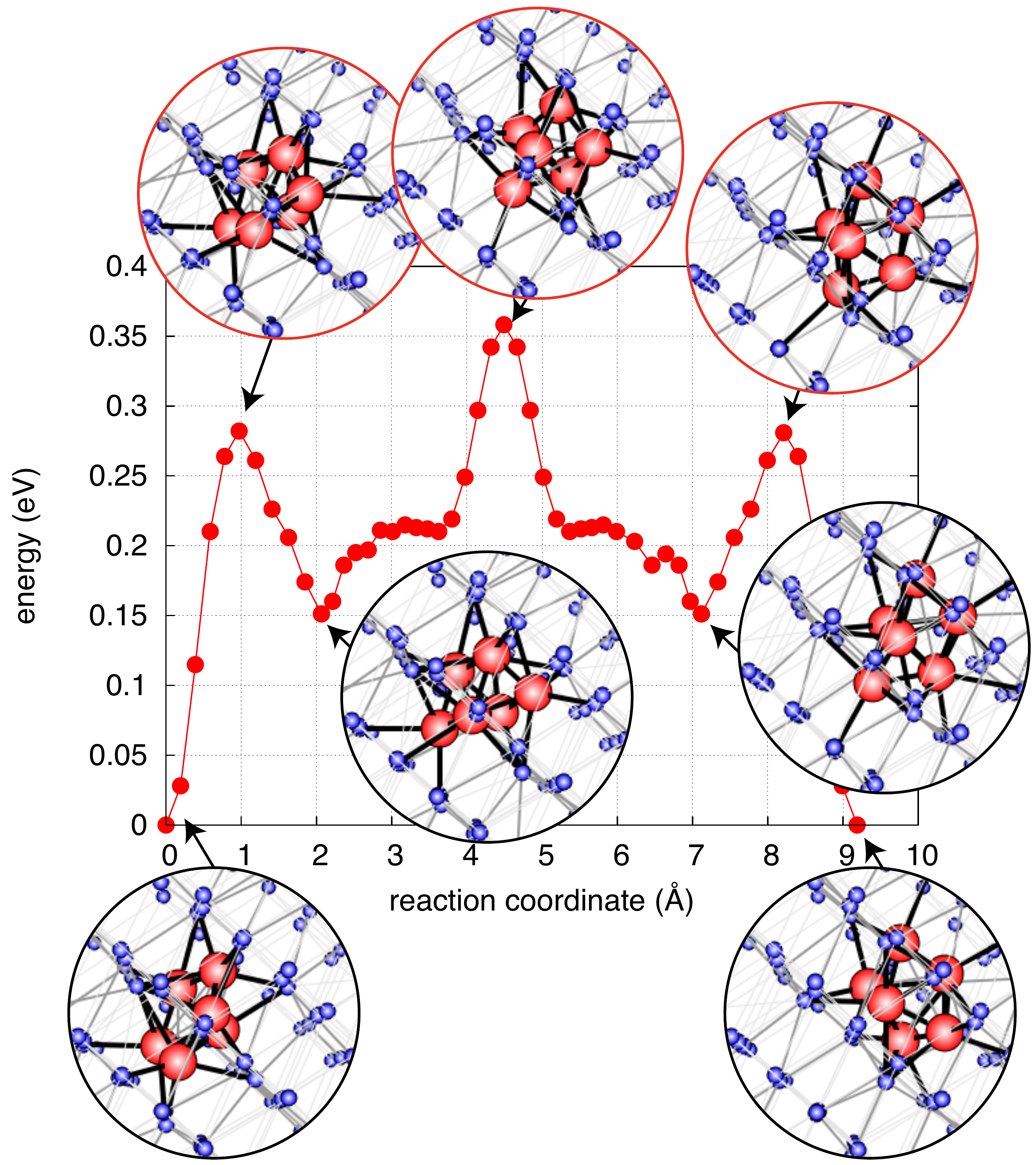}
\caption{Lowest energy migration pathway for interstitial He clusters
  from sizes $N=1$ to 6 as identified from TAD simulations. In these
  figures, the minimum energy path (MEP) is given by the points on the
  curve. Tungsten atoms are smaller/blue spheres while He atoms are
  larger/red spheres.}
    \label{fig:pathway}
\end{figure*}

Figure~\ref{fig:pathway} shows the lowest energy pathway found for
each cluster size. The minimum energy pathway, as identified from the
TAD simulations using the NEB method, is given as a function of the
reaction coordinate in angstroms. It should be noted that the NEBs
were not fully converged all along the path, only the highest energy
saddle point for each section of the path was fully converged.
However, the rest of the images are close to converged. In several
cases, the pathway involves multiple steps that were identified in the
TAD simulation as separate events. The pathways in
Fig.~\ref{fig:pathway} are a composite of the unique state-to-state
events that lead to the migration of the cluster.

For $N=1$, the simplest ``cluster'', the diffusion pathway goes from
the lowest energy structure, which is the He occupying a tetrahedral
interstice in the BCC lattice, through an octahedral interstice, which
is the saddle point, to another tetrahedral site. The barrier for this
process is 0.15\,eV. This process is essentially the opposite of that
found for carbon diffusing within BCC Fe~\cite{Simonovic2010}.

As the cluster size increases, the pathways for diffusion become more
complex, with more intermediate minima along pathways that describe
net translation of the cluster. For example, for $N=2$, there is an
intermediate minimum halfway through the diffusion process. This
intermediate minimum is relatively deep in energy, only 0.06\,eV
higher than the lowest energy state. For even larger cluster sizes,
there tend to be very shallow minima, or shoulders, on the side of the
largest saddle peaks in the minimum energy path. Many of these shallow
minima are associated with saddle points that are extremely small,
only a few meV in some cases (e.g. $N=6$). However, along each pathway
there are also relatively deep intermediate minima, sometimes almost
degenerate in energy with the lowest energy state (e.g. $N=3$).

\begin{figure}[t]
 \centering
 \includegraphics[width=6.7cm,trim=0.5cm 0.6cm 1.0cm 0.6cm,clip]{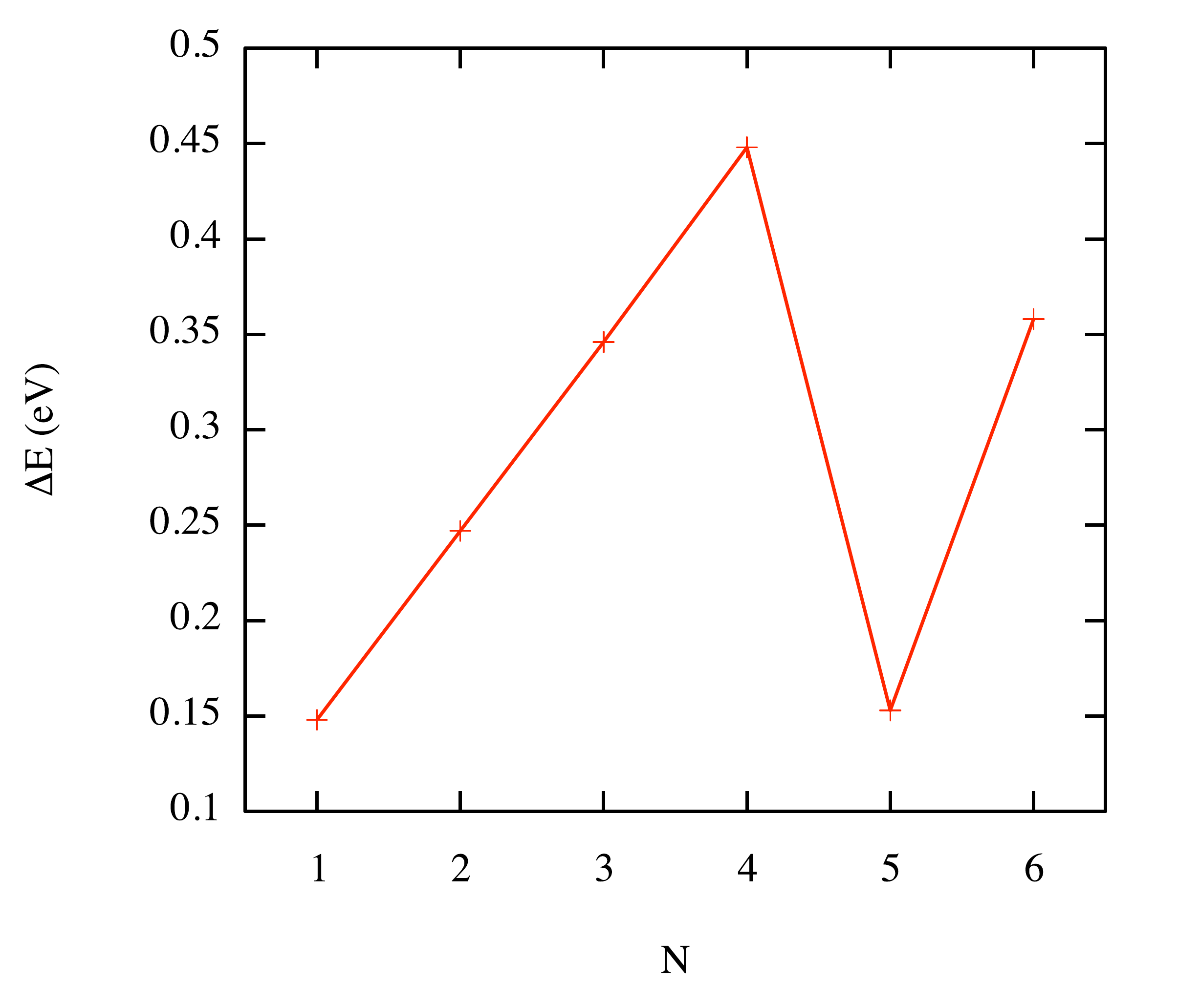}
    \caption{Migration energies as a function of cluster size for $N=1$ to 6.}    
    \label{fig:migration}
\end{figure} 
Results are summarized in Fig.~\ref{fig:migration}, in which the
energy of the highest energy saddle along the lowest energy pathway is
given versus the size of the cluster. Initially, as the size of the
cluster increases, the barrier increases linearly with the size of the
cluster. However, clusters of size 5 exhibit extremely high mobility,
which is again reduced significantly for $N=6$. This coincides with
the structure of the clusters. Recall that all of the clusters are
comprised of He atoms residing on tetrahedral sites with the lone
exception of $N=5$. That cluster has one He atom in an octahedral
interstice. Interestingly, similar anamolously fast diffusion for
He$_5$ interstitial clusters has been observed in
Fe~\cite{Stewart2011}. For $N=1$, the octahedral interstice is the
saddle point for migration, indicating that He at the octahedral
interstice is less energetically favorable than at the tetrahedral
site. This further implies that He mobility might be higher for such a
structure, which is reflected in the low migration energy.

\subsubsection{Diffusivity}

\begin{figure*}[]
 \includegraphics[width=.33\textwidth]{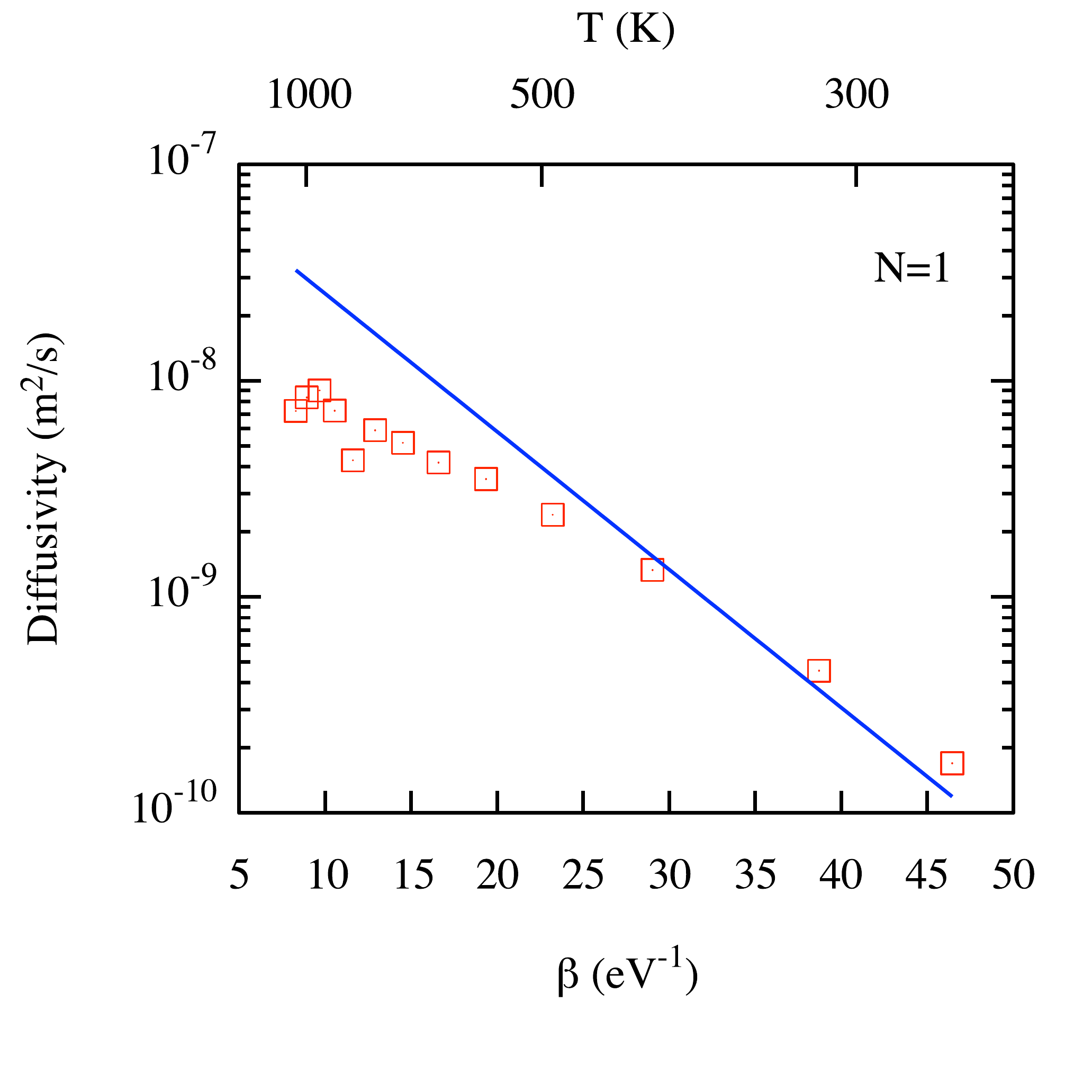}\hfill
 \includegraphics[width=.33\textwidth]{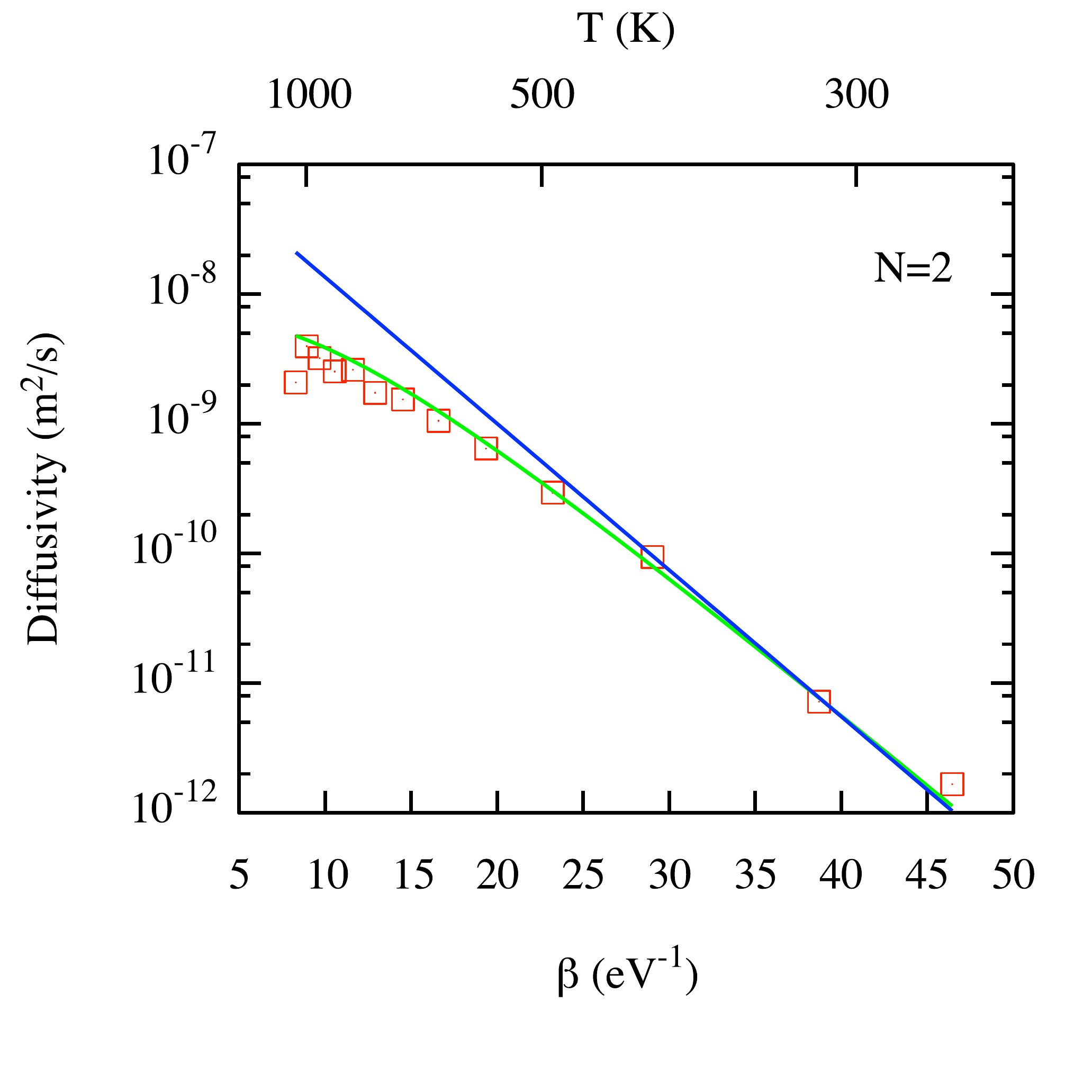}\hfill
\includegraphics[width=.33\textwidth]{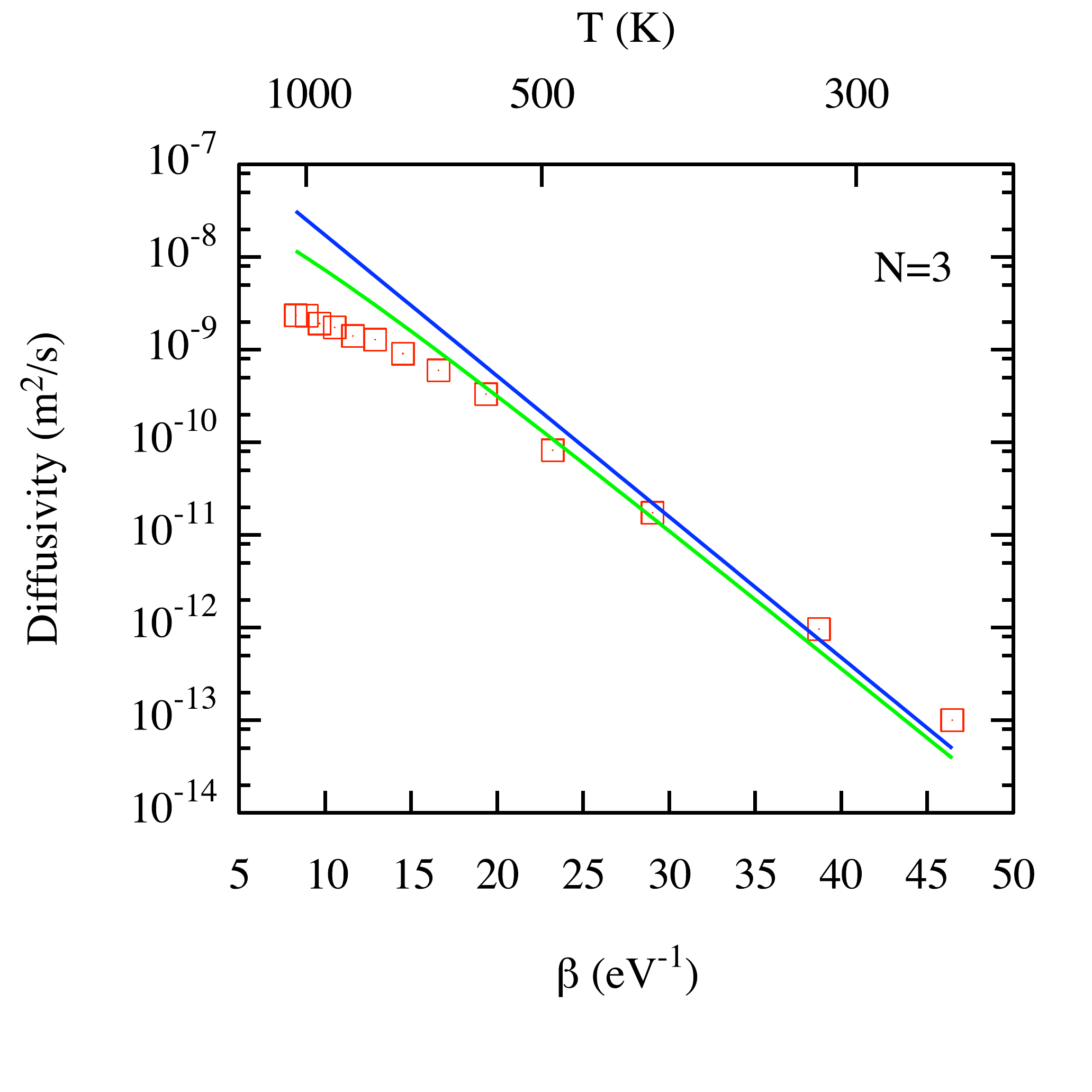}\\
 \includegraphics[width=.33\textwidth]{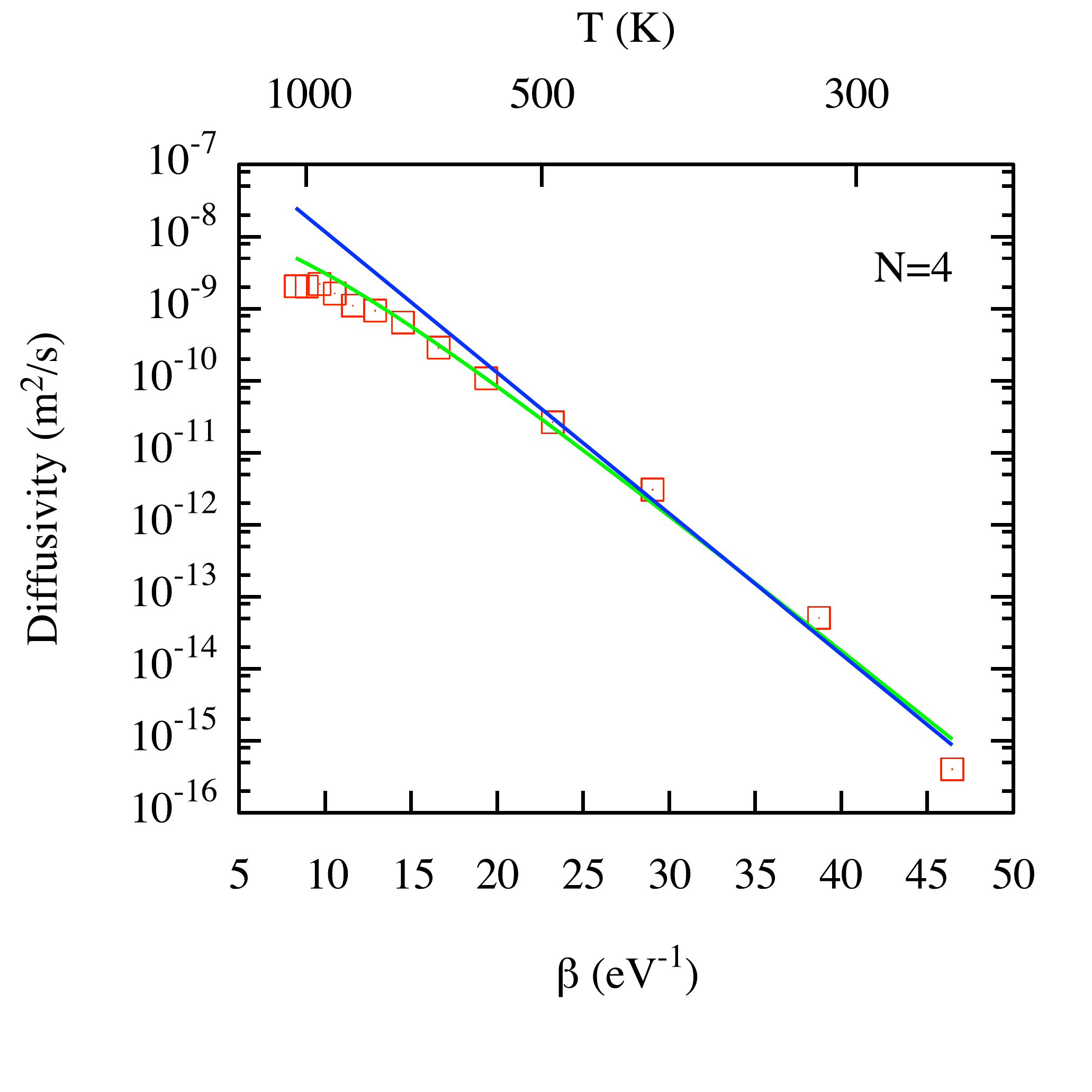}\hfill
 \includegraphics[width=.33\textwidth]{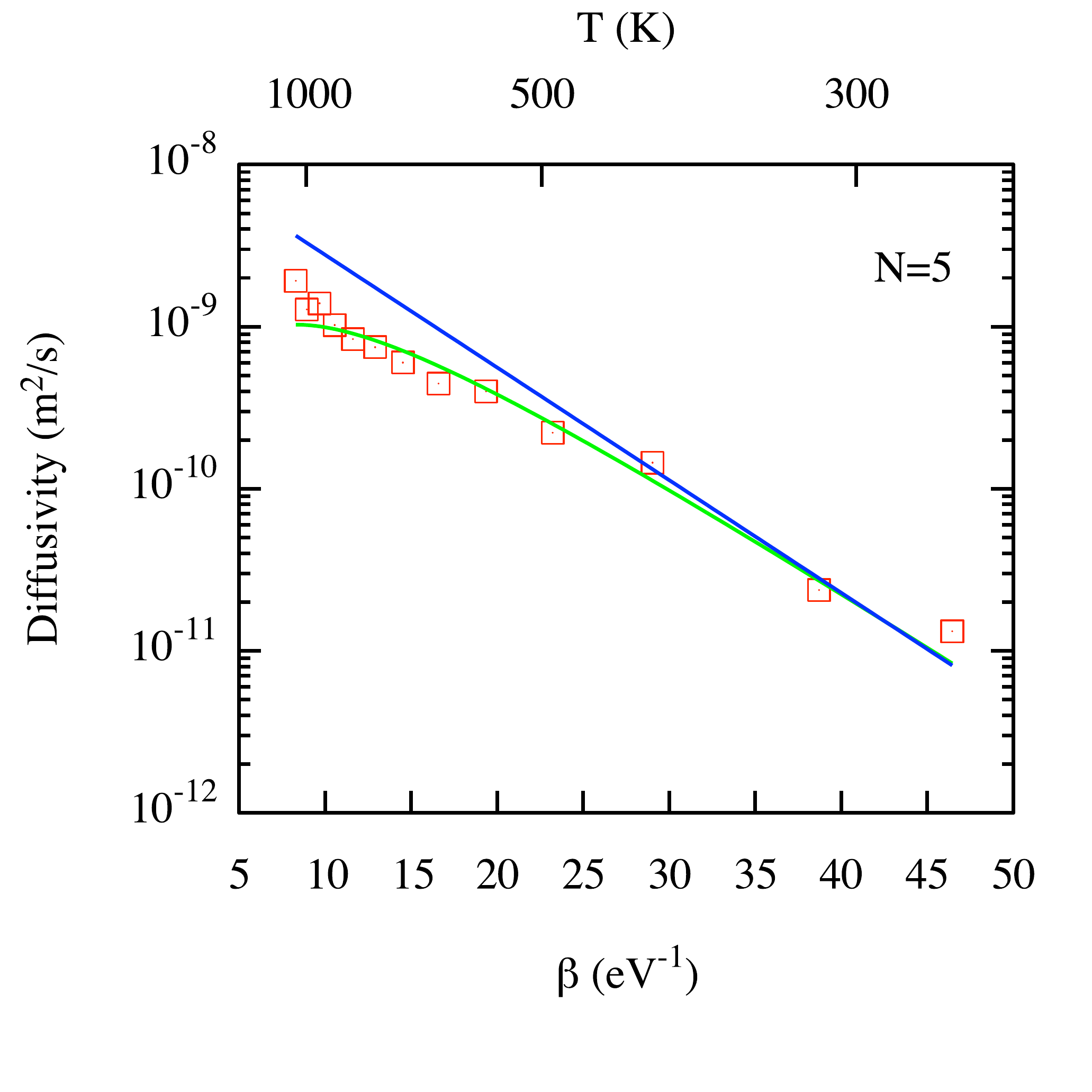}\hfill
\includegraphics[width=.33\textwidth]{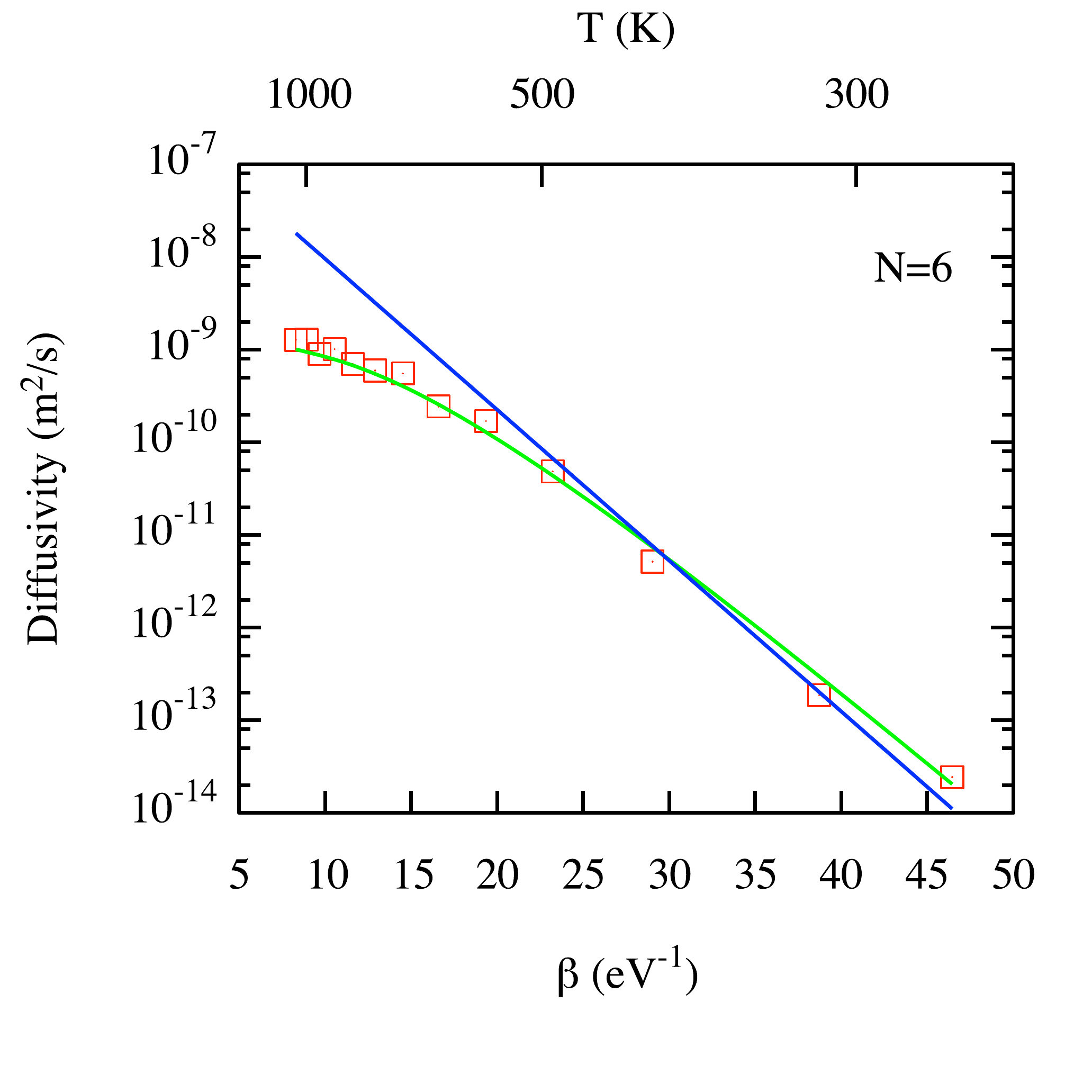}
    \caption{Diffusivity as a function of $\beta$ for $N=1$ to 6. Red squares: MD results; blue line: HTST; green line: SB-HTST.}    
    \label{fig:diffusivity}
\end{figure*}

The diffusivity of the different clusters was directly measured over
temperatures ranging from 300\,K to 1400\,K by
computing the mean squared displacement (MSD) of the cluster's center
of mass and using Einstein's relation
\begin{equation}
D= \lim_{t\rightarrow\infty}\frac{\langle |r(t) -r(0)|^2\rangle }{2dt},
\label{eq:einstein}
\end{equation}
where $d$ is the dimension, i.e., 3. In practice, we used 24
independent simulations, totalling about 600\,ns of MD time at each
temperature. The diffusivity was obtained from a linear fit to the
time-dependent MSD. The results are summarized in Fig.\
\ref{fig:diffusivity}. At first glance, the most obvious feature is
the departure from a conventional Arrhenius [$D\propto \exp(-\Delta
E/k_\textrm{B} T)$] behavior. While, at modest temperatures, the
change of the diffusivity is compatible with the diffusion barriers
identified with TAD (the blue lines are Arrhenius fits using the TAD
values of the activation energies), the Arrhenius curves bend
downwards at high temperatures, i.e., diffusion occurs slower than
suggested by an extrapolation from lower temperatures. This
observations is counter-intuitive because one could have expected
that, due to the complexity of the clusters' energy landscape, other,
higher energy diffusions pathways would have become active with
increasing temperature. This would have lead to an opposite, i.e.,
convex, deviation from the Arrhenius behavior.

The origin of this unexpected behavior indeed stems from the
complexity of the energy landscape. However, not in terms of a
multiplicity of diffusion pathways but of possible conformations of
the clusters.  Consider a case where only one transition pathway is
active but where the cluster has to be found in a specific
configuration for the hop to be possible. As the temperature varies,
the relative probability that the cluster be found in such a
``gateway'' conformation also varies, which leads to a non-Arrhenius
behavior, i.e., to a temperature-dependent effective activation
energy. In fact, an explicit expression for this energy can be derived
within the purview of what we term Super-Basin Transition State Theory
(SB-HTST):
\begin{equation}
\frac{\partial \ln k/k_0}{\partial \beta}=-\left[  U^*_G - \langle U_{\mathrm{min}} \rangle_{S,\beta} \right ] = -\Delta \tilde{E}(\beta)\,.
\label{eq-SBHTST}
\end{equation}
This expression states that the apparent activation energy (i.e., the
slope of the Arrhenius curve) corresponds to the difference between
the saddle point energy for diffusion $U^*_G $ and the average energy
of the minima (the different conformations) that the trajectory visits $\langle U_{\mathrm{min}}
\rangle_{S,\beta}$ during dynamics at inverse temperature $\beta$.
The complete derivation is presented in Annex \ref{annex:sbhtst}.
Finally, integrating this expression with respect to $\beta$, one can obtain a generalized
Arrhenius expression:
\begin{equation}
D =\hat{\nu} \exp(-\beta \Delta \hat{E})\,,
\end{equation}
where the generalized activation energy $\Delta \hat{E}$ is now a
function of temperature, but where the prefactor $\hat{\nu}$ is
temperature independent.

As can be seen from Eq.~(\ref{eq-SBHTST}), SB-HTST predicts a lowering
of the slope of the Arrhenius curve at increasingly high temperature,
due to the fact that higher-lying energy basins are sampled
increasingly often, in qualitative agreement with the MD results.
Except for a slight under-correction at $N=3$, SB-HTST in fact
quantitatively predicts the behavior of the diffusivity, as shown by
the green lines in Fig.~\ref{fig:diffusivity}. In this case, $\langle
U_{\mathrm{min}} \rangle_{S,\beta}$ is measured directly from MD
simulations by periodically quenching the trajectory and recording the
energy of the minima where it was instantaneously located. $\langle
U_{\mathrm{min}} \rangle_{S,\beta}$ was then fitted to a third order
polynomial in $T$, from which $\Delta \hat{E}=f_0+f_2 T^2 + f_3 T^3$
was finally obtained. The value of $\hat{\nu}$ was here fitted
directly to the MD data. The results are presented in
Table~\ref{tab:diffusivity}. For $N=3$, where the disagreement is the
largest, we verified that the MSD is directly proportional to the
number of crossings of the diffusion barrier identified by TAD over
the whole temperature range, i.e., no other pathway significantly
contributes to diffusion. As SB-HTST is an harmonic theory (i.e., it
assumes that the partition functions of each conformation can be
approximated by a many-dimensional harmonic partition function), the
most likely source of error is anharmonicity of the potential energy
surface. Interestingly, for $N=1$, we observe a similar departure from
a perfectly Arrhenius behavior. We did not observe any other
thermally-relevant configuration of the interstitial that could cause
a super-basin correction. Therefore, anharmonicity is probably also
the cause of the departure from the Arrhenius behavior in this case.
\begin{table}
\begin{ruledtabular}
\begin{tabular}{|c||c|c|c|c|}
  N & $\hat{\nu}$  (m$^2$ s$^{-1}$) & $f_0$ (eV) & $f_2$ (eV/K$^2$) & $f_3$ (eV/K$^3$)\\
\hline
\hline
  2 & $3.26 \times 10^{-7}$ & 0.26 &  $1.80 \times 10^{-7}$ & $-3.79
  \times 10^{-11}$\\
\hline
  3 & $5.61 \times 10^{-7}$ & 0.35 &  $8.02 \times 10^{-8}$ & $-1.48
  \times 10^{-11}$\\
\hline
 4 & $2.86 \times 10^{-6}$ & 0.46 &  $1.98 \times 10^{-7}$ & $-3.84
  \times 10^{-11}$\\
\hline
5 & $1.98 \times 10^{-8}$ & 0.16 &  $1.18 \times 10^{-7}$ & $-1.30
  \times 10^{-11}$\\
\hline
6 & $2.32 \times 10^{-6}$ & 0.375 &  $4.21 \times 10^{-7}$ & $-9.77
  \times 10^{-11}$\\
\hline

\end{tabular}
\end{ruledtabular}
\caption{\label{tab:diffusivity}Parameters of the SB-HTST description of
  the diffusivity for different cluster sizes. See text for details.}
\end{table}

\subsection{Breakup}
\label{subsec:breakup}

A second key process that controls the size distribution of He
clusters is the breakup of larger clusters into smaller ones. Even if
clustering is energetically favorable, configurational entropy
considerations favor isolated He atoms at low concentrations and high
temperatures. Proper consideration of breakup reactions is also
essential to model the time and position at which clusters will reach
a sufficient size to trap mutate and create a bubble nuclei.

\begin{figure}[]
    \centering
    \includegraphics[width=6.5 cm,trim=0.5cm 0.6cm 0.3cm 0.6cm,clip]{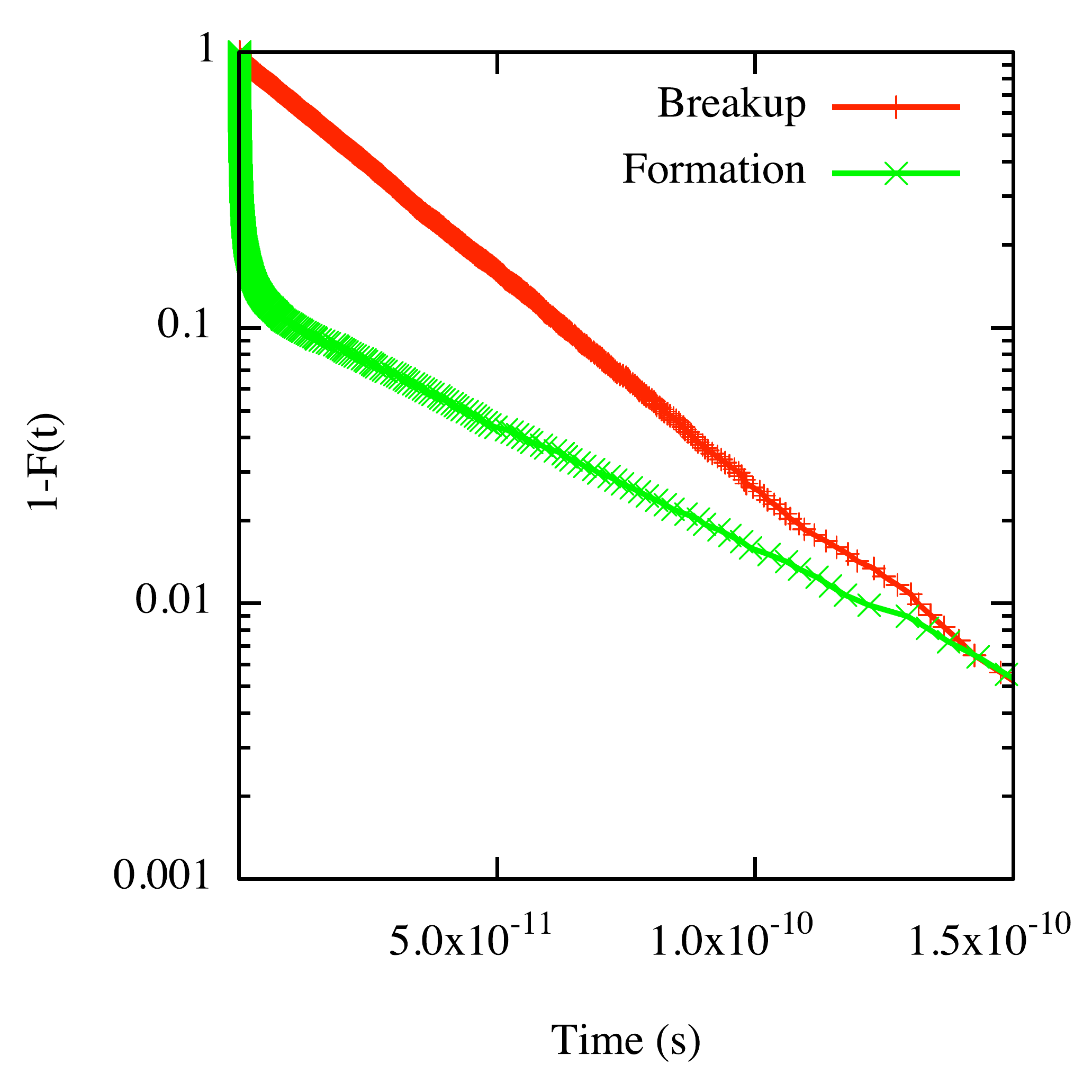}
    \caption{1 minus the empirical cumulative reaction time
      distribution function $F(t)$ for $N=2$ at $T=1500$\,K with
      $R_{\mathrm{max}}=2.8$\,\AA. Red: breakup; Green: formation.}
    \label{fig:reaction-time-1}
\end{figure}

\begin{figure}[]
    \centering
    \includegraphics[width=6.5 cm,trim=0.5cm 0.6cm 0.3cm 0.6cm,clip]{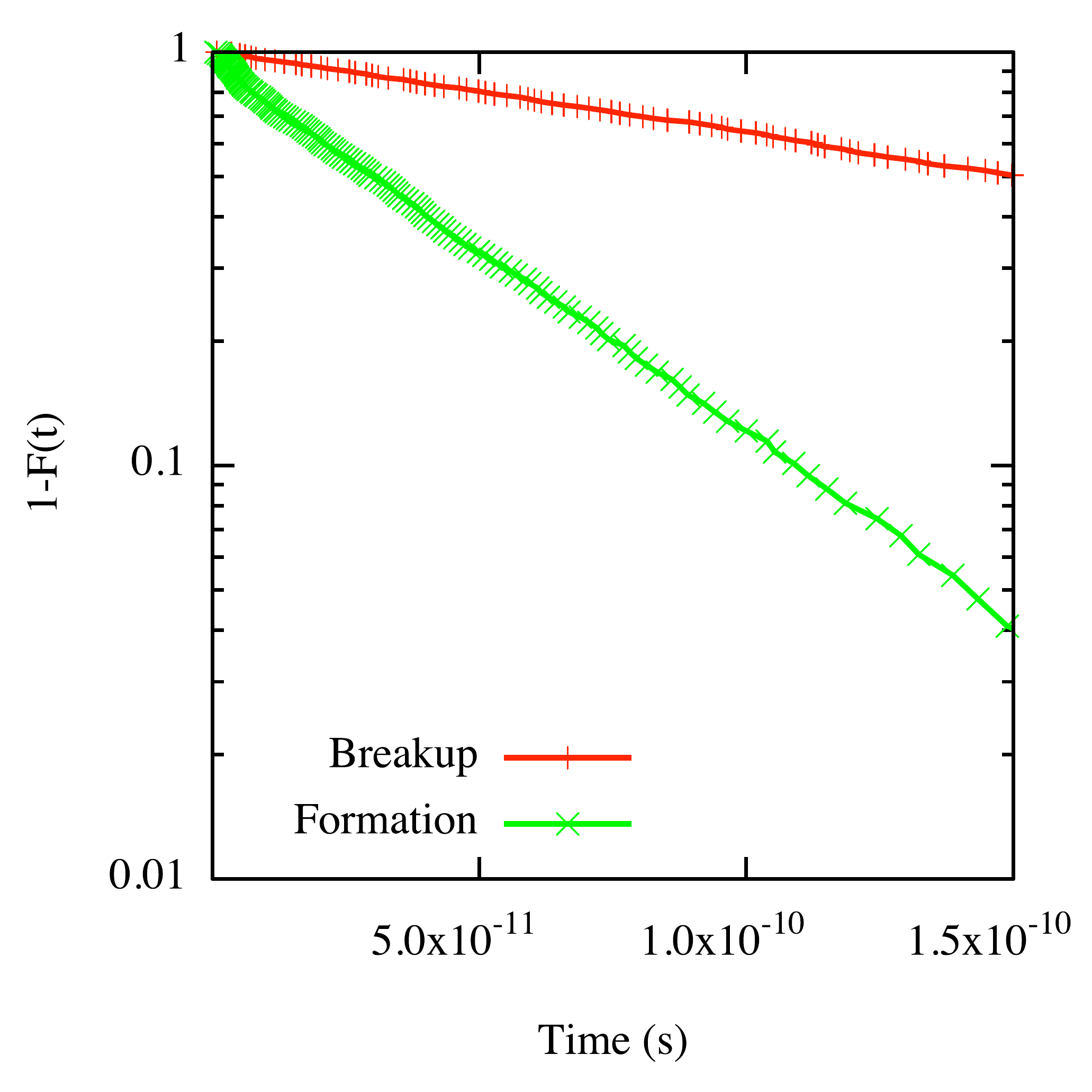}
    \caption{1 minus the empirical cumulative reaction time
      distribution function $F(t)$ for $N=2$ at $T=1500$\,K with
      $R_{\mathrm{bound}}=2$\,{\AA} and $R_{\mathrm{unbound}}=7$\,\AA.
      Red: breakup; Green: formation.}
    \label{fig:reaction-time-2}
\end{figure}

A proper definition of a suitable reaction coordinate to describe the
breakup process is essential. Unfortunately, a simple choice based on
a cutoff distance between atoms is not suitable. Indeed, defining
breakup (formation) as was done in the {thermodynamic} analysis above
--- as the moment at which the minimal distance $R$ within which every
member of the cluster can be connected to some other member of the
cluster exceeds (falls below) a certain threshold $R_{\mathrm{max}}$
--- gives an inadequate {\em kinetic} description. Indeed, an ideal
definition should be such that both breakup and (re-)formation of the
cluster is approximatively a Markovian process, i.e., that it be well
described by a rate constant (in fact, this condition is necessary,
e.g., if the dynamics are to be described by a rate theory or a
cluster dynamics model). Markovianity in turns implies that the
breakup and formation time distribution should be exponential, or,
equivalently, that the 1 minus the cummulative reaction time
distribution be exponential. The measured distributions for $N=2$ and
$T=1500$\,K for $R_{\mathrm{max}}=2.8$\AA, reported in
Fig.~\ref{fig:reaction-time-1}, show that this simple procedure is
inadequate because it is associated with an excess of rapid
(re-)formation of the cluster, i.e., many ``breakups'' really
correspond to a short-lived fluctuation of the cluster. As
$R_{\mathrm{max}}$ is increased, an excess of short-time
pseudo-formation also occur, corresponding to He atoms that came in
close proximity, without bonding for a significant amount of time.  We
found that no single value of $R_{\mathrm{max}}$ gives satisfactory
results. Alternatively, one might consider, following
Ref.~\onlinecite{vanden-eijnden2005}, that the fully bound
$R<R_{\mathrm{bound}}$ and fully unbound $R>R_{\mathrm{unbound}}$
regions of configuration space are well separated ($R_{\mathrm{bound}}
\neq R_{\mathrm{unbound}}$), i.e., that a cluster can temporarily be
found in neither the fully bound or fully unbound regions of
configuration space, but that is bound/unbound status can still be
determined by the last region it visited. In other words, a cluster is
bound if it more recently met the condition $R<R_{\mathrm{bound}}$
than $R>R_{\mathrm{unbound}}$, and vice-versa. In this formalism,
spurious rapid recrossings of the $R_{\mathrm{max}}$ surface are not
deemed reactions and only unambiguous breakups/formations are counted.
The reaction dynamics can therefore be made Markovian to a good
approximation, as shown in Fig.~\ref{fig:reaction-time-2} for
$R_{\mathrm{bound}}=2$\,{\AA} and $R_{\mathrm{unbound}}=7$\,\AA. We find
this definition to be adequate for all cluster sizes and temperatures
investigated here.

\begin{figure}[]
    \centering
    \includegraphics[width=6.5 cm,trim=0.5cm 1.6cm 0.4cm 0.1cm,clip]{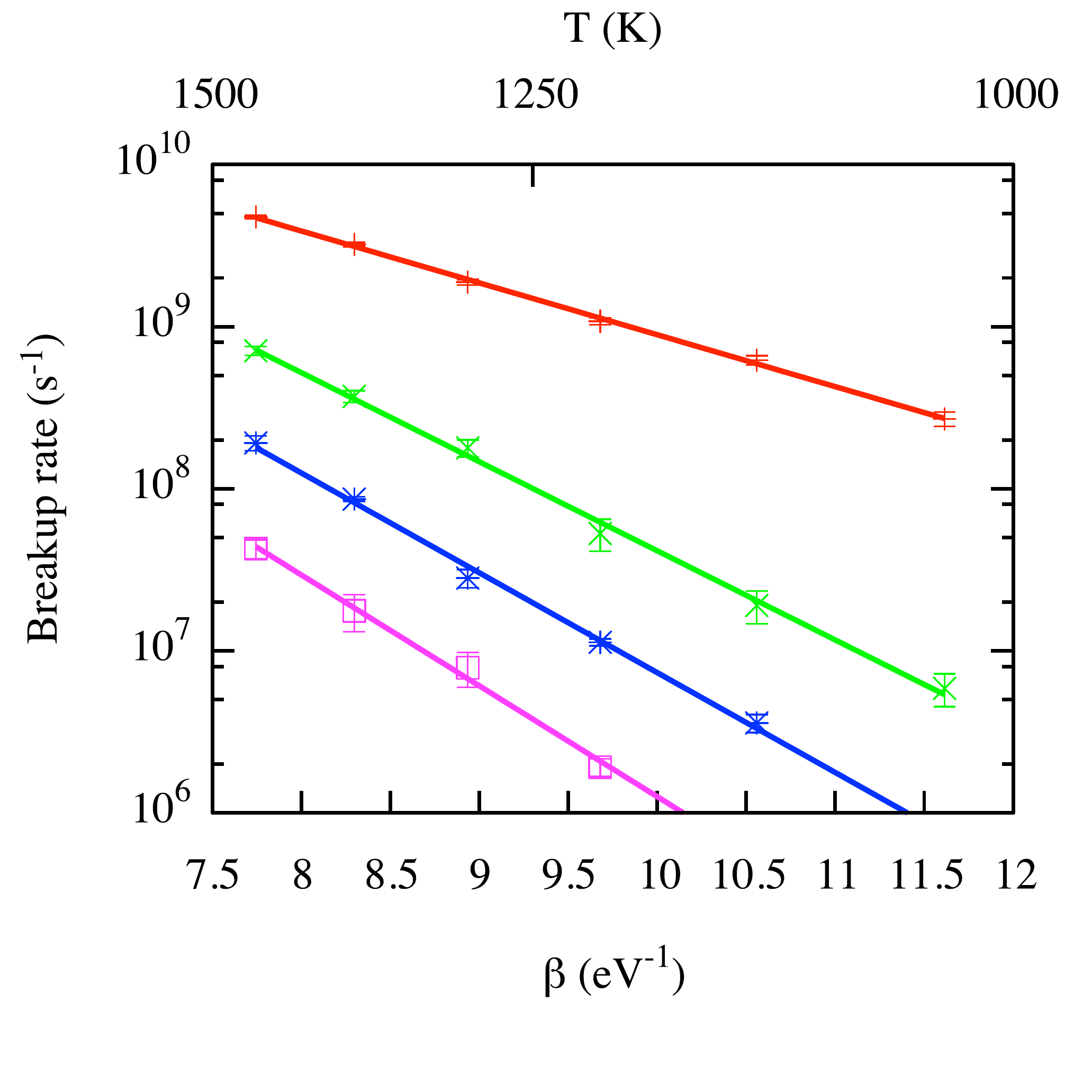}
    \caption{Breakup rate as a function of $\beta$. Red crosses:
      $N=2$; Green $\times$: $N=3$; Blue stars: $N=4$; Pink squares:
      $N=5$. Corresponding lines are Arrhenius fits.}
    \label{fig:breakup-rate}
\end{figure}

Using this approach, we directly measured the breakup rate of clusters
in MD between 1000 and 1500\,K. The breakup rates for $N=2$ to 5 are
shown in Fig.~\ref{fig:breakup-rate}. At 1000\,K, the lifetime of
clusters is on the order of 10\,ns, 0.1\,$\mu$s, 1\,$\mu$s, and
10\,$\mu$s, for $N=2$, 3, 4, and 5, respectively.  We did not observe
breakup for $N=6$ on accessible timescales. The results indicate a
rapid decrease of the breakup rate with increasing cluster size. As
shown in Table \ref{tab:breakup}, where the results of Arrhenius fits
to the breakup rates are reported, the activation barriers for breakup
increase significantly with increasing size, which consistent with the
fact that the binding energy per He atom increases with increasing
cluster size. In fact, these activation barriers are similar to the
binding energy differences for the removal of single He from clusters
(cf., Fig.\ \ref{fig:energies}).  The corresponding prefactors are
fairly standard, i.e., around $10^{12}$ to $10^{13}$\,s$^{-1}$.

Given the very large number of transition pathways (the reverse of
every possible clustering pathway), we did not attempt to exhaustively
identify them all in order to perform an SB-HTST correction (as the
net effect would result from a competition between the super-basin
correction and the activation of more pathways).  The observed breakup
rates are remarkably Arrhenius over the probed temperature range, but
deviations at lower temperatures cannot be fully excluded.  

While we expect the number of distinct breakup pathways to be large,
the vast majority of the breakups involve a single atom leaving the
cluster. At elevated temperatures more complex transitions start to
activate, but their contribution is modest: at 1500\,K, the
$5\rightarrow 3 + 2$ transition occurs only about 4\% of the time, the
$4\rightarrow 2 + 2$ about 2\% of the time. That fraction becomes even
more modest at lower temperatures.

% While we did not attempt to fully characterize the available breakup
% pathways, the analysis of the post-breakup configurations indeed
% confirm that more than one variant is active. Branching ratios for
% different product states are summarized in Tables
% \ref{tab:breakup-branch-3} and \ref{tab:breakup-branch-4} for $N=3$, and
% 4. In both cases, the most likely scenario is the detachment of a
% single He atom from the cluster, but other, more complex transitions
% --- such as the decomposition of a cluster into fragments --- are also
% active, especially at elevated temperatures. This is consistent with
% the observation that binding becomes stronger as the cluster size
% increases. 
\begin{table}
\begin{ruledtabular}
\begin{tabular}{|l||l|l|}
  N & $\nu$  (s$^{-1}$) & $\Delta E$ (eV)\\
\hline
\hline
  2 & $1.41 \times 10^{12}$ & 0.74  \\
\hline
3 & $1.29 \times 10^{13}$ & 1.27 \\
\hline
4 & $1.05 \times 10^{13}$ & 1.41 \\
\hline
5 & $8.66 \times 10^{12}$ & 1.57 \\
\end{tabular}
\end{ruledtabular}
\caption{\label{tab:breakup}Prefactors and energy barriers for cluster breakup obtained from an Arrhenius fit to the MD data.}
\end{table}

\subsection{Trap Mutation}
\label{subsec:mutation}

The last process of interest to microstructural evolution is the
so-called ``trap mutation'' process, whereby an interstitial cluster
of He induces the creation of a W Frenkel pair and collapses into the
newly created vacancy. From then on, the cluster is practically
immobile and can be considered a nano-bubble. Additional He atoms
encountering it will also join the bubble, which will grow by further
emitting W interstitials~\cite{Greenwood1959}. We postpone the
analysis of the growth process to a future publication; instead we
here assess the rate at which clusters can undergo this mutation. The
mutation process was observed in direct MD simulations.

\begin{figure}[]
    \centering
a)\\
\includegraphics[width=5.5 cm]{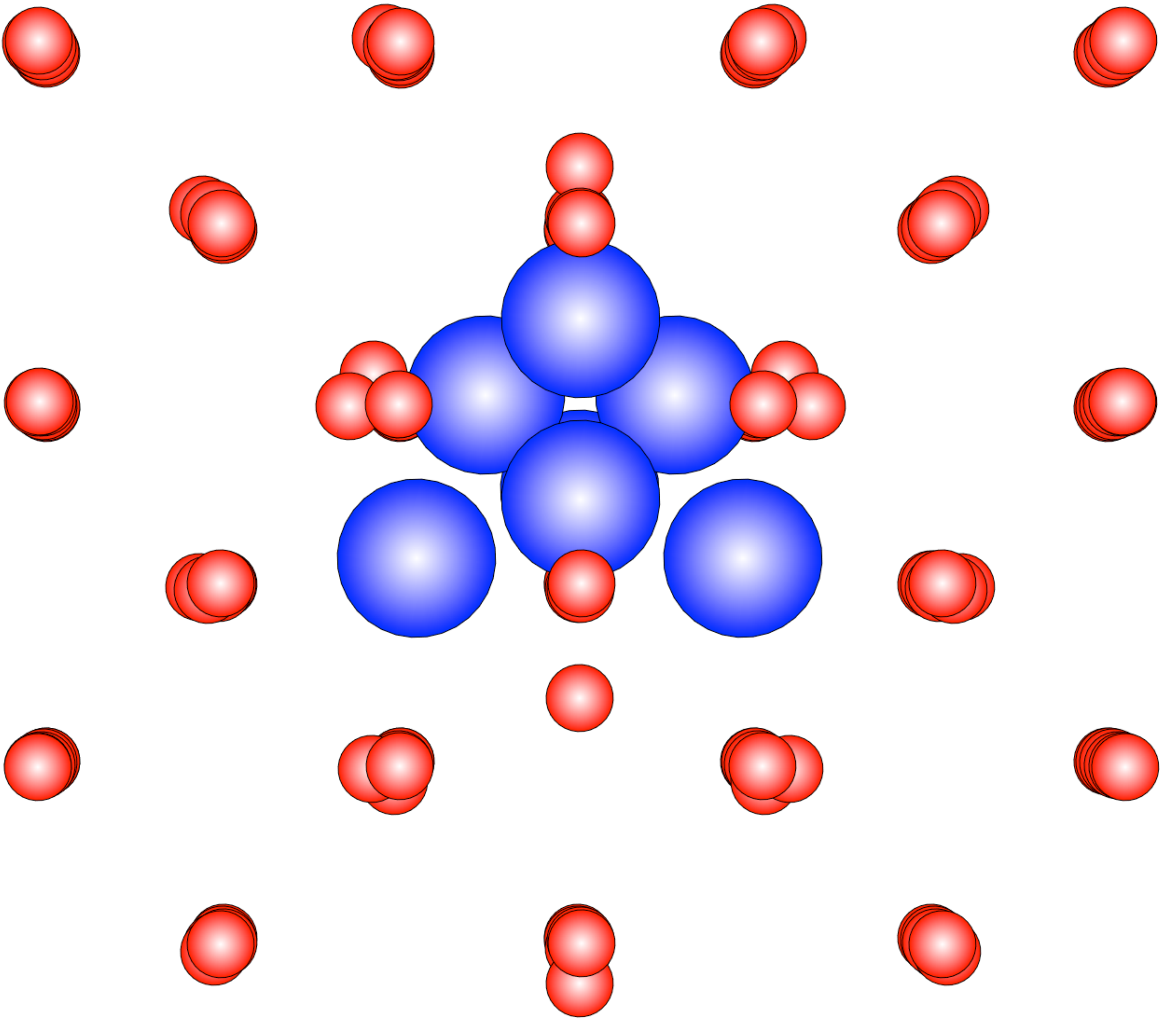}\vspace{0.5cm}\\
b)\\
\includegraphics[width=5.5 cm]{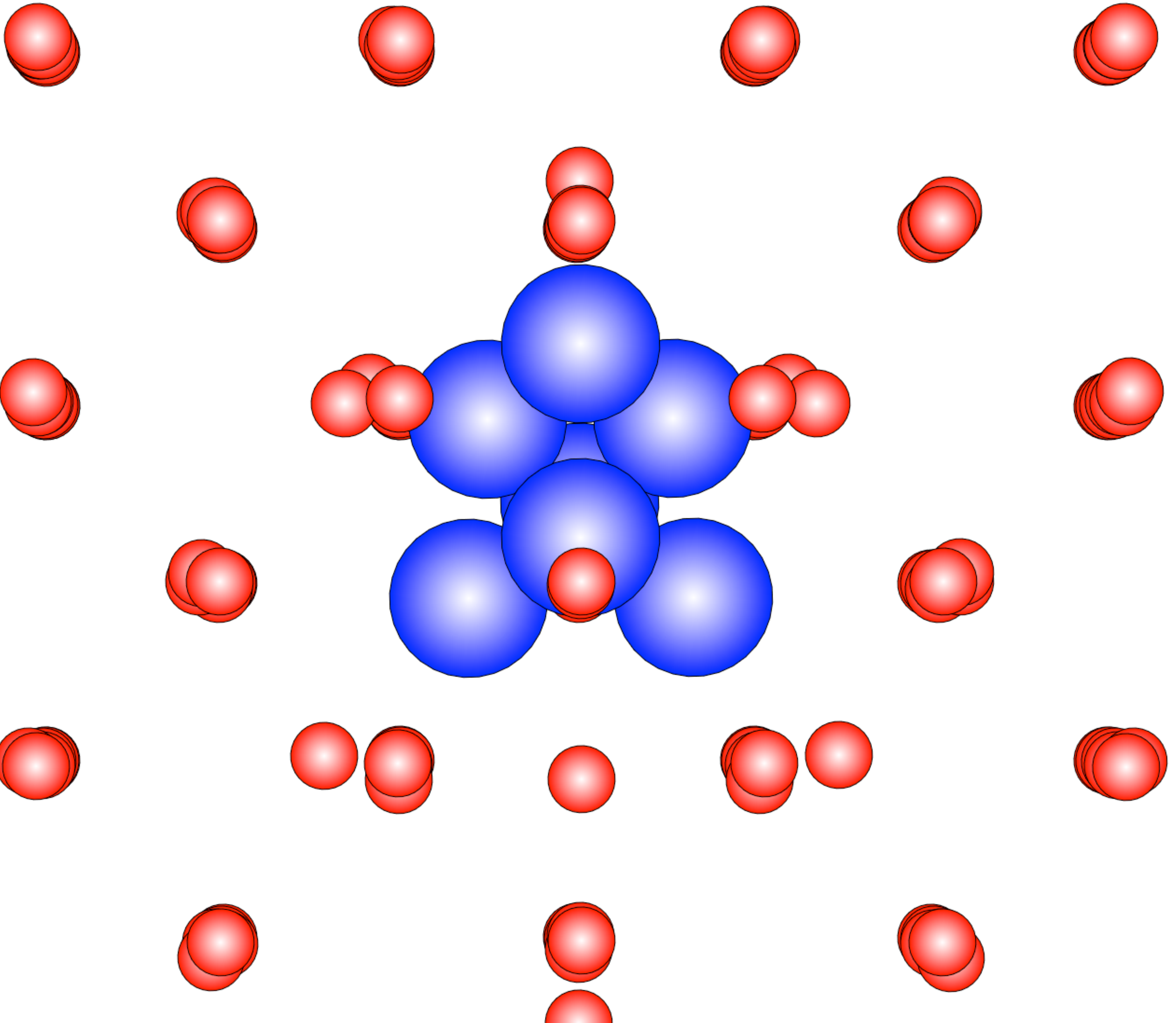}\vspace{0.5cm}\\
c)\\
\includegraphics[width=5.5 cm]{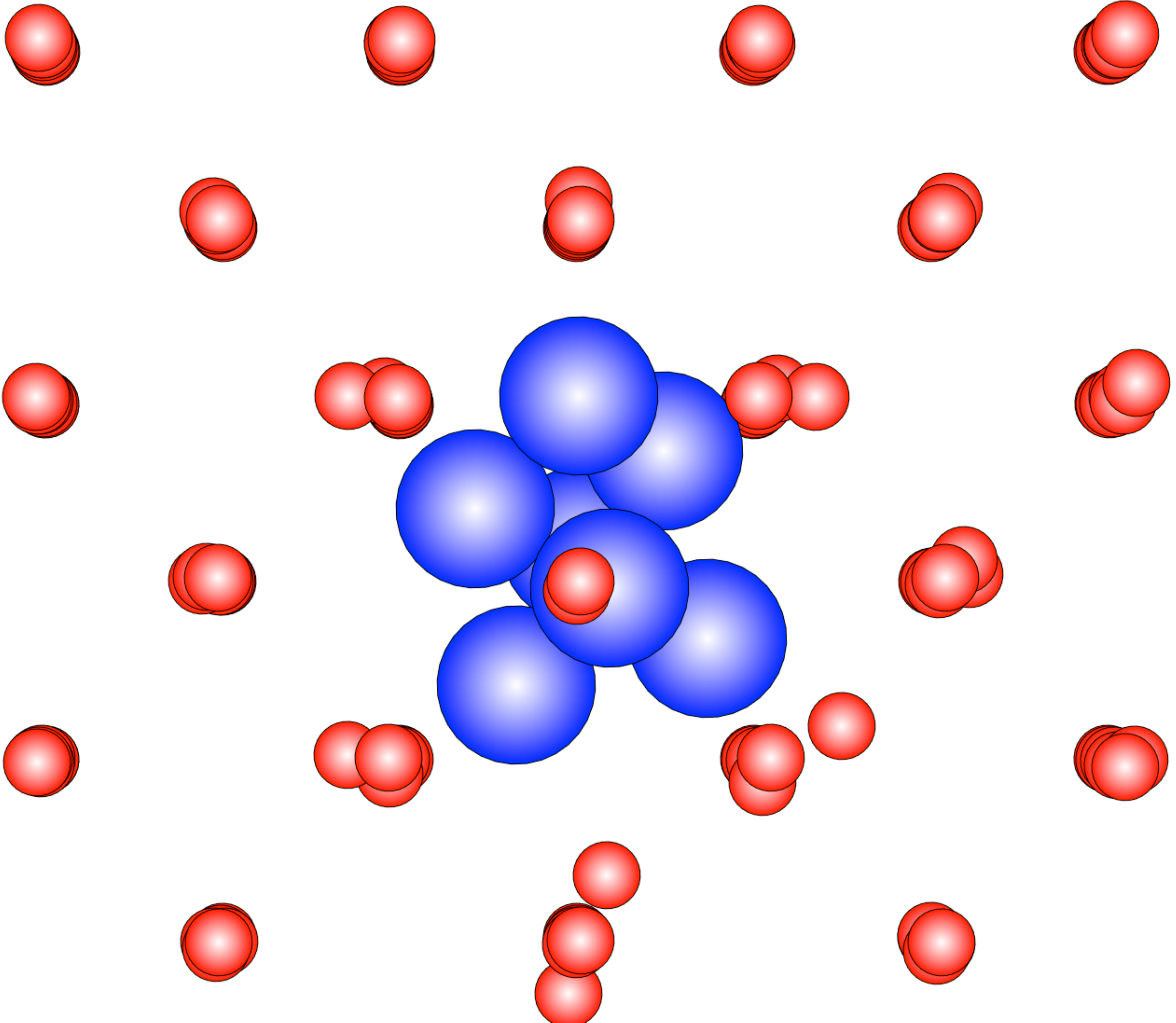}
    \caption{Possible trap mutation pathway of an He$_7$ cluster. The three
      snapshots correspond to the initial (a), saddle (b), and
      final (c) configurations. }    
    \label{fig:mutation-pathway}
\end{figure}

A typical mutation event is illustrated in
Fig.~\ref{fig:mutation-pathway}.  This process, with an energy barrier
of 1.06\,eV, was identified from an MD trajectory at 1400\,K and
characterized using the NEB method. In the initial state, one can
clearly see that a W atom was found especially far from its
equilibrium configuration [cf., panel a), under the cluster]. As the
cluster becomes more compact [cf., panel b)], this atom is further
pushed away, which finally leads to its complete ejection [cf., panel
c)] and to the formation of an interstitial. After a few
rearrangments, the interstitial rearranges into a crowdion
configuration (not shown). This process admits many variants (we have
identified a selection of these), but the associated barriers appear
to be similar to the one described above.

\begin{figure}[]
    \centering
    \includegraphics[width=6.5 cm,trim=0.5cm 1.6cm 0.6cm 0.1cm,clip]{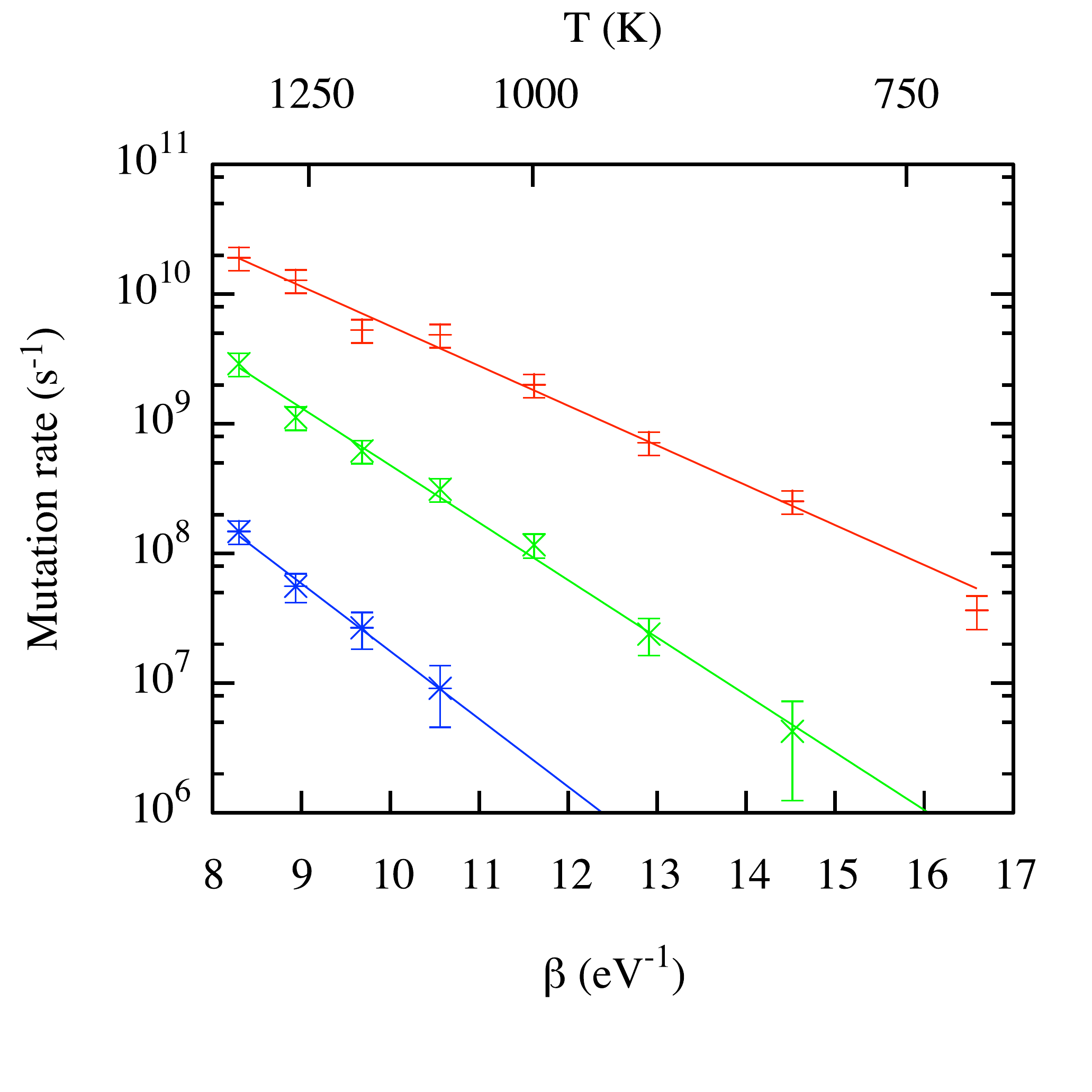}
    \caption{Mutation rate as a function of $\beta$. Red crosses:
      $N=7$; Green $\times$: $N=6$; Blue stars: $N=5$. Corresponding
      lines are Arrhenius fits.}
    \label{fig:mutation-rate}
\end{figure}

As shown in Fig.~\ref{fig:mutation-rate}, the mutation rate increases
sharply with size, as expected. At 1000\,K, the lifetime of a cluster
before mutation varies from a few hundred ns for $N=7$ down to a
fraction of a $\mu$s for $N=5$. Once again, the behavior of the rate
is nicely Arrhenius over the range of temperature we investigated
(700--1400\,K). The results of Arrhenius fits to the MD results are
summarized in Table~\ref{tab:mutation}.  The energy barriers extracted
from the fits, ranging from 0.701\,eV for $N=7$ to 1.20\,eV for $N=5$,
are significantly lower than that of the reaction pathways we found
(e.g., $1.06$\,eV for the $N=7$ transition shown in
Fig.~\ref{fig:mutation-pathway}), but the disagreement decreases
significantly upon consideration of SB-HTST corrections. For example,
for $N=7$, the MD results are consistent with a raw barrier of about
0.9 eV. Based on the slight tendency for under-correction observed in
the case of diffusion, it is plausible that a process with a barrier
of about 1 eV, as the one described above, fully accounts for the
observed results. However, the possibility that lower barrier pathways
also contributes to the rate cannot be excluded.
\begin{table}
\begin{ruledtabular}
\begin{tabular}{|l||l|l|}
  N & $\nu$  (s$^{-1}$) & $\Delta E$ (eV)\\
\hline
\hline
  5 & $2.95 \times 10^{12}$ & 1.20 \\
\hline
6 & $1.28 \times 10^{13}$ & 1.02 \\
\hline
7 & $6.64 \times 10^{12}$ & 0.701 \\
\end{tabular}
\end{ruledtabular}
\caption{\label{tab:mutation}Prefactors and energy barriers for cluster mutation obtained from an Arrhenius fit to the MD data.}
\end{table}

Interestingly, the inverse reaction (the re-absorption of an
interstitial) was also observed for $N=5$ and 6. The rate for this
process is however difficult to determine precisely because the
relatively small size of the simulation cell used here might bias the
results, as the created interstitial cannot be ejected far away from
the cluster due to periodic boundary conditions. Therefore, it
remains confined in the vicinity, which artificially increases the
recombination rate. We however observe that mutation makes lower
energy states accessible for $N=7$, but not for $N=5$, and 6. This is
consistent with a higher rate of recombination for smaller clusters.

\section{Discussion}
\label{sec:discussion}

\begin{figure}[]
    \centering
    \includegraphics[width=6.5 cm,trim=0.5cm 0.6cm 0.5cm 0.6cm,clip]{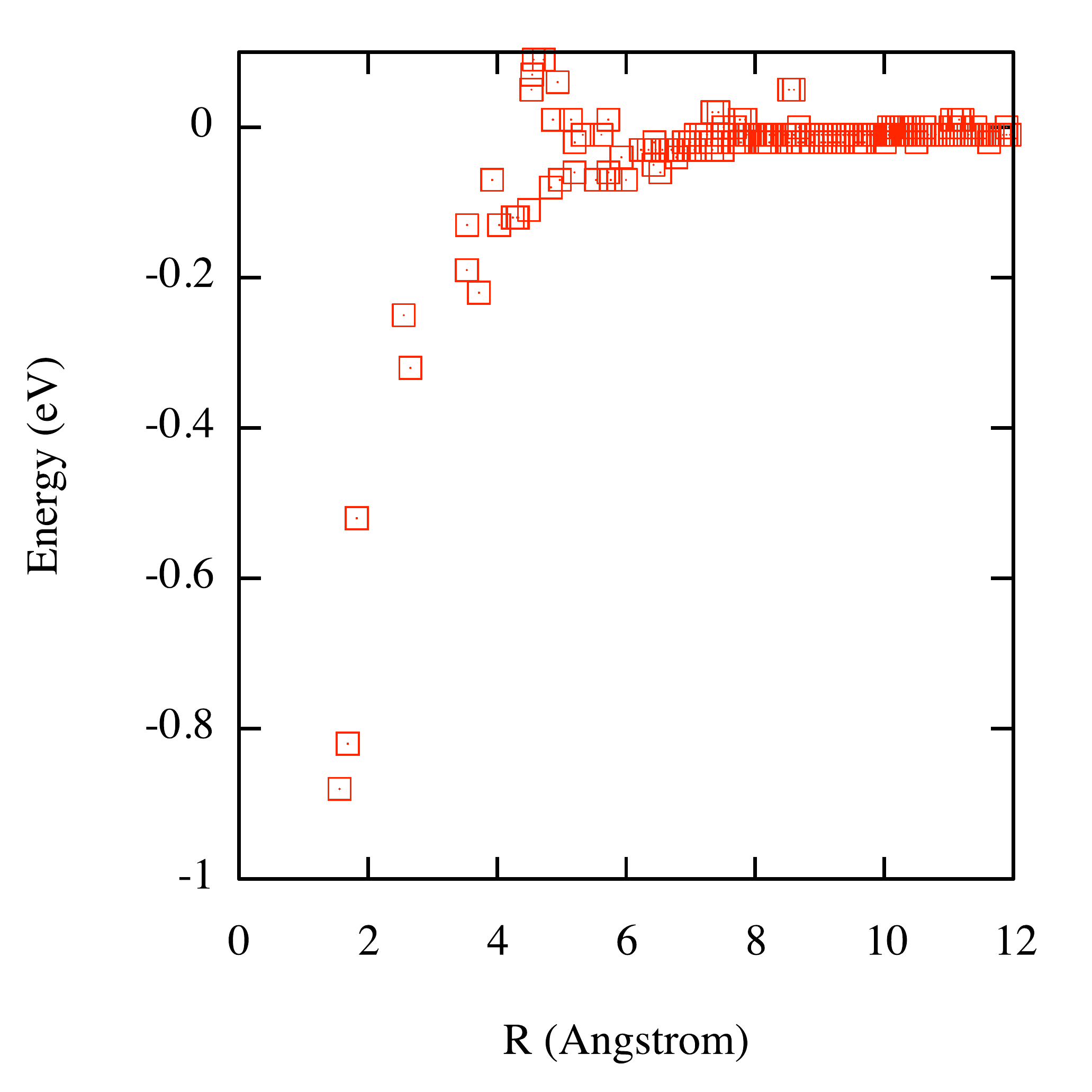}
    \caption{Potential energy of quenched configurations taken along a
      2000\,K MD simulation as a function of the distance between 2 He
      atoms.}
    \label{fig:ener-r-2}
\end{figure}

The quantities reported above (free energies, diffusivities, breakup
rates, and mutation rates) can be used to upscale atomistic simulation
through mesoscale cluster dynamics models where the population of
individual species is obtained through the solution of a set of
coupled reaction/diffusion equations~\cite{Marian2012}. These models
can be used to bridge the gap between the nanoscale and relevant
device scales and provide information on the microstructural evolution
of the material over long timescale and large lengthscales. In that
context, the reaction rates between different species are commonly
obtained by assuming diffusion-limited reactions, i.e., that the
formation rate constant of a specie $C$ by collision and merger of two
other species $A$ and $B$ (He clusters of different sizes in the
present context) adopts a Smoluchowski form~\cite{Smoluchowski}:
\begin{equation}
k_{A+B\rightarrow C}=4\pi(D_A+D_B)(r_A+r_B),
\label{eq:smoluchowski}
\end{equation}
where the $D$ are diffusivities and the $r$ effective capture radii.
Using the measured formation rates (of the form
$k_{\mathrm{He}_N+\mathrm{He}\rightarrow \mathrm{He}_{N+1}}$), we can
extract the relevant capture radii from the atomistic results. We
obtain $r_N\simeq$ 4.3, 9.9, 15, and 25\,{\AA} around 1000\,K, for
$N=1$, 2, 3, and 4, respectively, decreasing to about $r_N\simeq$ 2.9,
7.3, 9.2, and 13\,{\AA} around 1500\,K. These values are surprisingly
large compared to an effective hard core radius of a single He atom.
This points to a significant contribution from elastic interactions
mediated by the tungsten lattice. This is qualitatively consistent
with the observation that He atoms bind increasingly strongly with
increasing cluster size, even in absence of purely chemical binding
between He atoms. This hypothesis can be confirmed directly by
calculating the energy of different configurations as a function of
the distance between 2 He atoms. For clarity, configurations taken
along a trajectory at 2000\,K have been quenched before computing the
distance and potential energy. As shown in Fig.~\ref{fig:ener-r-2},
the elastic interaction between a pair of He atoms extends up to a
range of about 6\,\AA. Postulating that capture occurs once the
interaction energy reaches about $k_\mathrm{B} T$ yields results that are in
reasonable agreement with the radii infered from~Eq.~(\ref{eq:smoluchowski}).

A commonly taken approach when parameterizing cluster dynamics models
is to rely on Eq.~(\ref{eq:smoluchowski}) combined with a detailed
balance condition to estimate the breakup rates of the clusters. This
approach is attractive in cases where free energy differences are
computable using sophisticated sampling techniques and diffusivities
and capture radii can be estimated using MD simulations, but where
breakup rates are too low to be directly measured. However, as
discussed in Sec.~\ref{subsec:breakup}, a purely thermodynamical
description of the breakup process is here inadequate because it leads
to non-Markovian breakup kinetics. We next assess whether using the
kinetic approach to computing the formation rates and the
thermodynamic approach to compute the binding free energy yields
accurate results.

\begin{figure}[]
    \centering
    \includegraphics[width=6.7 cm,trim=0.5cm 1.6cm 0.3cm 0.1cm,clip]{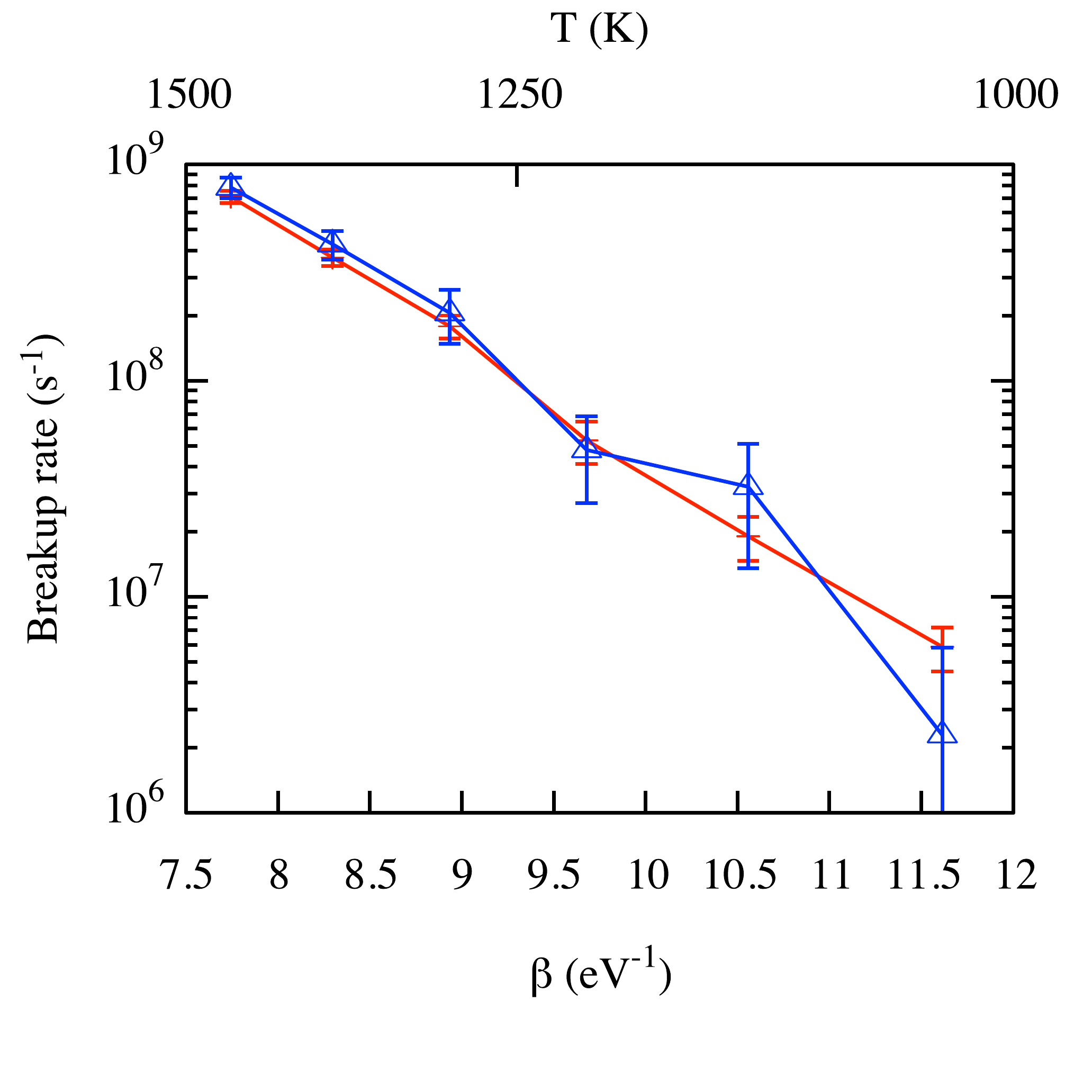}
    \caption{Breakup rate for the reaction He$_3\rightarrow$
      He$_2$+He. Red: direct MD simulations (cf. Fig.
      \ref{fig:breakup-rate}); blue: from Eq.~(\ref{eq:breakup-FE}).  }
    \label{fig:breakup-comparison}
\end{figure}

Using the free energy change upon breakup $\Delta F_{C\rightarrow A+B}$, one gets:
\begin{equation}
k_{C\rightarrow A+B} =\frac{1}{V} k_{A+B\rightarrow C} \exp ( -\beta \Delta F_{C\rightarrow A+B} ).
\label{eq:breakup-FE}
\end{equation} 
Using this last equation with the measured formation rates [or,
equivalently, using the infered (temperature-dependent) capture radii]
and the calculated binding free energies, the accuracy of this
approach can be assessed by comparing to the directly measured breakup
rates. As shown in Fig.~\ref{fig:breakup-comparison} for the breakup
of He$_3$, these two approaches are in close agreement (similar
agreement is observed for other reactions). This suggests that, while
the absolute values of the formation and breakup rates significantly
differ in the thermodynamic and kinetic formalism, their ratio is
approximatively the same, i.e., that the correction to the rate can be
described using a simple transmission factor.

Our results have some interesting consequences regarding the behavior
of He in tungsten. Consider the $N=5$ cluster. At $T=1000$\,K, it has
a diffusivity of 9.14$\times 10^{-10}$\,m$^2$s$^{-1}$, a rate to
break-up of 1.06$\times 10^{5}$\,s$^{-1}$, and a rate to trap mutate of
2.64$\times 10^{6}$\,s$^{-1}$. Thus, the rate to trap mutate is about 20
times that for breaking up at 1000\,K and, on the time scale of trap
mutation, the cluster can diffuse about 19\,nm. In contrast, at 500\,K,
trap mutation is nearly 2000 times more likely than break-up and the
cluster would diffuse over 10\,$\mu$m before trap mutating. Thus,
whether a cluster such as $N=5$ would contribute more to the growth or
nucleation of bubbles would be very temperature dependent. This has
important consequences for the behavior predicted in higher level
models and illustrates the need to obtain accurate rates for all
relevant processes.

The behavior of these clusters is clearly not a simple function of
their size, with particularly the $N=5$ cluster exhibiting what might be
deemed anomalous behavior as compared with the other clusters. The
behavior of the $N=5$ cluster is correlated with its structure which, among
the clusters examined here, is the sole cluster to contain a He atom
in an octahedral position in its ground state. Thus, there is some
relationship between, for example, the structure of the cluster and
its kinetic properties, though that relationship is not trivial.

Finally, it is interesting to note that, even though these clusters,
especially the larger ones, exhibit a very rich and complex landscape
of local minima (hundreds of different conformations), their migration
rates are very well described by considering them as exisiting in a
super basin that has a key escape pathway. This leads to an excellent
description of the migration rates as a function of temperature using
SB-HTST. These rates are still harmonic on a basin-to-basin level, but
anharmonicity is effectively captured to first order through the fact
that the average energy of the minimum is itself a function of
temperature. We expect that this modification of harmonic TST will
provide a powerful avenue for analyzing the rates of complex
structures such as the clusters described here. It should alse be very
useful in any situation where large regions of configuration spaces
are confined by only a few kinetic bottlenecks. In these cases, we
would expect the super-basin corrections to be significant. While a
brute-force approach to the parameterization of all the inter-basin
rates would be prohibitively expensive, the escape rates out of these
bottlenecks can be readily obtained with SB-HTST.

\section{Conclusions}
We investigated the kinetics and thermodynamics of small He clusters
in W in conditions relevant to fusion energy production.  Our
simulations yield insights into the structure and thermodynamics of
these clusters, and provide a complete characterization of their
kinetics in terms of diffusion, breakup, and mutation into nanoscale
He bubbles. Our results enable the parameterization of cluster
dynamics models that can bridge the gap between the nano and
mesoscales and hence facilitate the prediction of the performance of W
as a first-wall material in the next generation of fusion reactors.

\begin{acknowledgments}

  This work was supported by the United States Department of Energy
  (US DOE) SciDAC program.  Los Alamos National Laboratory is operated
  by Los Alamos National Security, LLC, for the National Nuclear
  Security administration of the US DOE under contract
  DE-AC52-06NA25396. This research used resources of the National
  Energy Research Scientific Computing Center, which is supported by
  the Office of Science of the U.S. Department of Energy under
  Contract No. DE-AC02-05CH11231.

\end{acknowledgments}

\appendix

\section{Super-Basin Transition State Theory}
\label{annex:sbhtst}
Conventional harmonic transition state theory (HTST)~\cite{Vineyard}
is usually appropriate when the initial (reactant) state is composed
of a single basin of attraction of the potential energy surface. We
now show that it can be generalized in cases where the initial state
is a super-basin, i.e., a collection of individual basins. Individual
basins can represent different conformations of the same defect, and
inter-super-basins transitions can correspond to a net motion of the
defect, or to a certain reaction.  Assume that the transition path of
interest originates in basin $G$ and leaves the super-basin $S$
through a dividing surface $G^*$. The TST rate $k$ at which this
transition occurs is simply given by the probability of finding the
system in $G$ relative to elsewhere in the super-basin multiplied by
the TST rate for escape from $G$ across $G^*$, i.e.:
\begin{equation}
k=p_G k_{G\rightarrow G^*}=\frac{Z_G}{Z_{S}} \frac{1}{\beta h} \frac{Z^*_G}{Z_G}=\frac{1}{\beta h} \frac{Z^*_G}{Z_{S}},
\label{eq:tst}
\end{equation}
where the $Z$ are canonical partition functions. 

Computing the slope of the Arrhenius curve at a given inverse
temperature $\beta$,
\begin{equation}
\frac{\partial \ln k/k_0}{\partial \beta}=\left. \frac{1}{k}\frac{\partial k}{\partial \beta}\right|_\beta,
\end{equation}
where $k_0$ is an arbitraty constant with the units of a rate.
Using Eq.~(\ref{eq:tst}):
\begin{equation}
\frac{\partial \ln k/k_0}{\partial \beta}=\left[ -\frac{1}{\beta}+\frac{1}{Z^*_G}\frac{\partial Z^*_G}{\partial \beta}-\frac{1}{Z_{S}}\frac{\partial Z_{S}}{\partial \beta}\right].
\end{equation}
From the fact that $\partial Z / \partial \beta = \langle -E
\rangle_\beta Z$, with $E$ the total energy, we have:
\begin{equation}
\frac{\partial \ln k/k_0}{\partial \beta}=\left[ -\frac{1}{\beta}+-\langle E \rangle_{G^*,\beta} + \langle E \rangle_{S,\beta} \right].
\end{equation}

In the harmonic approximation, the equipartition theorem holds, and we get:
\begin{equation}
\frac{\partial \ln k/k_0}{\partial \beta}=-\left[  U^*_G - \langle U_{\mathrm{min}} \rangle_{S,\beta} \right ] = -\Delta \tilde{E}(\beta),
\label{eq-SBHTST-final}
\end{equation}
where $U^*_G$ is the potential energy at the saddle point in $G^*$ and
$\langle U_{\mathrm{min}} \rangle_{S,\beta}$ is the average potential
energy of the minimum of the basin the system is currently in. In
analogy with conventional Arrhenius kinetics, $\Delta
\tilde{E}(\beta)$ is interpreted as an effective, temperature
dependent, activation energy. As the temperature increases, the system
will spend increasing amounts of time in higher energy basins, which
will lead to a corresponding decrease of the effective activation
energy.  $\langle U_{\mathrm{min}} \rangle_{S,\beta}$, and hence
$\Delta \tilde{E}(\beta)$, can easily be obtained from direct MD
simulations, or estimated by computing harmonic partition functions.
Eq.~(\ref{eq-SBHTST-final}) can then be integrated to give the rate
$k(\beta)$.  Note that in extreme cases one could even observe a {\em
  negative} effective activation energy if some local minima in the
super-basin are located at energies higher than the saddle $U^*_G$.

%\bibliography{he-clusters}

%merlin.mbs 2010-03-15 4.21a (PWD, AO, DPC)

\end{document}